\documentclass[onecolumn,amsmath,amssymb,nofootinbib,letterpaper,11pt]{article}
\pdfoutput=1
\usepackage{jheppub}
\usepackage{ifpdf}
\usepackage[utf8]{inputenc}
\usepackage[T1]{fontenc}
\usepackage{graphicx}
\input{epsf}
\usepackage{epsfig}
\usepackage{epstopdf}
\usepackage{arcs}
\usepackage{subfigure}
\usepackage{tensor}
\usepackage{braket}
\usepackage{float}
\usepackage[dvipsnames]{xcolor}
\usepackage{amsmath}
\usepackage{morefloats}
\newcommand{\be}{\begin{equation}}
\newcommand{\ee}{\end{equation}}

\usepackage{tikz}
\usetikzlibrary{calc}   
\usetikzlibrary{topaths}
\usepackage{hyperref}
\hypersetup{
	colorlinks =WildStrawberry,
	linkcolor =red,
	pdftitle={Leading corrections to Kerr geometry},
	citecolor=NavyBlue,
	filecolor=WildStrawberry,
	urlcolor=WildStrawberry,
}

\usepackage[toc]{appendix}
\usepackage{color,soul} 
\usepackage{datetime}
\newcommand{\bea}{\begin{eqnarray}}
\newcommand{\eea}{\end{eqnarray}}
\newcommand{\ba}{\begin{eqnarray}}
\newcommand{\ea}{\end{eqnarray}}

\newcommand{\beq}{\begin{equation}}
\newcommand{\eeq}{\end{equation}}
\newcommand{\beqa}{\begin{eqnarray}}
\newcommand{\eeqa}{\end{eqnarray}}
\newcommand{\beqar}{\begin{eqnarray*}}
\newcommand{\eeqar}{\end{eqnarray*}}

\renewcommand{\href}[2]{#2}

\title{Thermodynamic geometry of the novel 4-D Gauss Bonnet AdS Black Hole}
\author[a]{Seyed Ali Hosseini Mansoori}
\affiliation[a]{Faculty of Physics, Shahrood University of Technology, P.O. Box 3619995161 Shahrood, Iran}
\emailAdd{ shosseini@shahroodut.ac.ir; shossein@ipm.ir}

\abstract{In this paper, the new formalism of thermodynamic geometry proposed in \cite{HosseiniMansoori:2019jcs} is employed in investigating phase transition points and the critical behavior of a Gauss Bonnet-AdS black hole in four dimensional spacetime. In this regard, extrinsic and intrinsic curvatures of a certain kind of hypersurface  immersed in the thermodynamic manifold contain information about stability/instability of heat capacities. We, therefore, calculate the intrinsic curvature of the $Q$-zero hypersurface for a four-dimensional neutral Gauss Bonnet black hole case in the extended phase space. Interestingly, intrinsic curvature can be positive for small black holes at low temperature, which indicates a repulsive interaction among black hole microstructures. This finding is in contrast with the five-dimensional neutral Gauss Bonnet black hole with only dominant attractive interaction between its microstructures.
 
}

\preprint{
}

\begin{document}
\maketitle

\section{Introduction}\label{a}


In recent years there has been growing interest in the study of the small-large black hole phase transition in AdS black holes and bearing some similarity to the liquid-gas phase transition of the Van der Waals (VdW) fluid \cite{Chamblin:1999tk, Chamblin:1999hg}. It has now been demonstrated that there is a complete identification between a charged AdS black hole and a VdW system  in the extended phase space where the cosmological constant treats as 
pressure and its conjugate quantity as volume \cite{Kubiznak:2012wp}. Several studies, for example \cite{Kubiznak,Altamirano,AltamiranoKubiznak,Altamirano3,Dolan,Liu,Wei2,Frassino,
Cai,XuZhao,Hennigar,Hennigar2,Tjoa2,Ruihong,WeiSci,Sherkatghanad:2014hda} have been performed on phase transitions of AdS black holes in the extended phase space. 

Several attempts have also been made to find out critical behavior of black holes by using Riemannian thermodynamic geometry \cite{reff8,reff9,reffff9,reff12,reff10,reff4,reff5,reff6,Zamani,reff14,Banerjee:2010da,Banerjee:2010bx,Banerjee:2016nse}.  In particular, the scalar curvature associated with such a geometry can provide us with useful information about black hole phase transitions. 
Preliminary work was carried out by Weinhold in early of 1970s \cite{reff8}. He introduced the phase thermodynamic space and developed a geometric description of the equilibrium space of a thermodynamic system. Motivated by this, Ruppeiner proposed a different metric structure for the equilibrium
space by using fluctuation theory of thermal states \cite{reff9,reff10}. More precisely, Weinhold's metric are made by a Hessian of the internal energy function, whereas Ruppeiner's metric elements are defined by the Hessian matrix of the entropy. It turns out that the line elements of Weinhold and Ruppeiner geometries are related to each other with the temperature $T$ as the conformal factor. As suggested in Ref. \cite{reff9}, there is a direct correspondence between singularities of the scalar curvature of the Ruppeiner geometry and phase transition points. However, in some contradictory examples \cite{reff5, Sarkar:2006tg}, Ruppeiner geometry fails to save this correspondence. In spite of this fact, Ruppeiner geometry could equip us with a powerful tool needed for understanding the microstructure of a thermodynamic systems \cite{reff9}. The line element of Ruppeiner geometry measures the distance between two equilibrium states in such a way that the probability of fluctuating between two neighboring states is larger than those for two distant states. 
For a given fluid system, Ruppeiner curvature can also be an indicator of microstructure interactions so that the positive/negative scalar curvature indicates a repulsive/attractive interaction dominates, while vanishing curvature implies that repulsive and attractive interactions are in balance \cite{reff9, reffff9}. 

To evade the issue of Ruppeiner gemometry in establishing a one-to-one correspodence between pahse transions and curvature singularities, we proposed a new formalism of Ruppeiner geometry, developed from considerations about thermodynamic potentials related to the mass (instead of the entropy) by Legendre transformations \cite{HosseiniMansoori:2019jcs, reff12, Mansoori:2016jer, reff13}. This allows us to find a one-to-one correspondence between  divergences of heat capacities and curvature singularities. Moreover, the well-known geometry like geometrothermodynamics (GTD) \cite{reff14} can be constructed by an explicit conformal transformation, which is singular at unphysical points were generated in GTD metric, from our geometry. 

In addition, as one takes free energy as thermodynamic potential and temperature and volume as the fluctuation coordinate in our formalism, the alternative form of Ruppeiner geometry introduced in Refs. \cite{ Wei:2019uqg,Wei:2019yvs} can be reproduced. 
 When this metric form was applied to the van der Waals fluid only a dominant attractive interaction was observed, while for a RN AdS black hole, in a small parameter range, a repulsive interaction was also found in addition to the dominant attractive interaction between black hole molecules. This approach has been extended to other black hole systems \cite{GW1,GW2,GW3,GW4,GW5,GW6,GW7}.

Recently, a novel Einstein-Gauss-Bonnet (EGB) garvity has been suggested in Ref. \cite{Glavan:2019inb} by starting from a $D$-dimensional spacetime, re-scaling the Gauss-Bonnet coupling $\alpha$ by a factor of $1/(D-4)$, and then taking the limit $D \to 4$.  As a result, the Gauss-Bonnet term gives rise to non-trivial contributions to gravitational dynamics in four dimensional spacetime \cite{Glavan:2019inb}. From the Lovelock's theorem, however, we know that Gauss-Bonnet term does not contribute to the equations of motion. Therefore, the model would violate at least one of the Lovelock conditions. As expected, some of the subtleties of this model appeared very soon \cite{Gurses:2020ofy, Ai:2020peo,Shu:2020cjw, Hennigar:2020lsl, Tian:2020nzb, Mahapatra:2020rds, Bonifacio:2020vbk}. For example, in \cite{Gurses:2020ofy} it was demonstrated that the novel 4D EGB gravity does not admit a description in terms of a covariant-conserved rank-2 tensor in four dimensions, because one part of the GB tensor always remains higher dimensional, while \cite{Hennigar:2020lsl, Tian:2020nzb} focused on more complicated solutions such as Taub-NUT solutions, showing the naive limit of the higher-dimensional theory to $D = 4$ is not well defined. However, these issues can be circumvented by considering regularized versions of 4D EGB gravity \cite{Lu:2020iav, Hennigar:2020lsl, Fernandes:2020nbq}. In Ref. \cite{Lu:2020iav}, the authors used a Kaluza-Klein-like procedure to generate a four-dimensional limit of Gauss-Bonnet gravity by compactifying $D$ dimensional EGB gravity on $D -4$  dimensional maximally symmetric space, followed by taking the limit where the dimension of this space vanishes. Moreover, in 
\cite{Hennigar:2020lsl, Fernandes:2020nbq} proposed a well defined 4D EGB gravity by generalizing a method employed by Mann and Ross to obtain a limit of the Einstein gravity in $D = 2$ dimensions \cite{Mann:1992ar}. The resulting theories possesses an additional scalar degree of freedom and are a special case of Horndeski theory \cite{Horen}. The recent paper \cite{Aoki:2020lig}, however, clarified the situation by taking the limit at the nonlinear level. They have shown that the limit either breaks a part of diffeomorphism or leads to extra degrees of freedom. This conclusion is in agreement with the Lovelock's theorem and also it is in agreement with the conclusions of the above mentioned papers. The authors in \cite{Aoki:2020lig} went further and formulated a consistent theory with two degrees of freedom by breaking the time diffeomorphism. They then have shown that the cosmological and black hole solution (with which we are interested in this paper) are also solutions of that consistent theory. The stability and shadow of this black hole and quasi-normal modes of a scalar, electromagnetic, and gravitational perturbations have been studied in \cite{Konoplya:2020bxa}. However, it is not clear whether the analysis of quasi-normal modes based on \cite{Glavan:2019inb} will be in agreement with the consistent theory defined in \cite{Aoki:2020lig}. For instance, in \cite{Arrechea:2020evj} it is shown that tensor perturbations at the second order of perturbations are ill-defined for \cite{Glavan:2019inb} while it will be not the case for the consistent model investigated in \cite{Aoki:2020lig} at the nonlinear level.

  An increasing number of studies have been carried out on novel 4D Einstein Gauss-Bonnet gravity. For example, the geodesic motions in the background of spherically symmetric black holes by focusing on the innermost stable circular orbits has been investigated in \cite{Back1}. 
The solutions of charged black hole \cite{Fernandes:2020rpa} and a rotating analogy of 4D GB black hole using Newman-Janis algorithm \cite{Wei:2020ght} have been also studied. In addition, the possible range of GB coupling parameter can be estimated by modeling the $M87^{*}$ as a rotating 4D GB black hole \cite{Wei:2020ght,Kumar:2020owy}. Thermodynamics of asymptotically AdS black hole solutions in the four dimensional EGB theory has been reported in \cite{Hegde:2020xlv,Singh:2020xju}, and it was observed that a VdW like phase transition exists.

 
In this paper, we attempt to study the critical behavior of 4D Einstein Gauss-Bonnet-AdS black holes around phase transition points by using our new formalism of thermodynamic geometry (NTG). This paper is organized as follows. Section \ref{sec2} gives a brief overview of the new formalism of thermodynamic geometry (NTG). We propose a new procedure for NTG geometry constructed from thermodynamic potentials, which generated by Legendre transformations, in order to establish a one-to-one correspondence between related curvature singularities and phase transitions. In Sections \ref{sec3} and \ref{sec4}, we apply NTG geometry to understanding the behavior of phase transition points of 4D charged GB black holes in the normal and extended phase space, respectively. Remarkably, extrinsic and intrinsic curvatures associated with NTG geometry reveal some information about the critical behavior and  microstructures of the black hole. Our conclusions are drawn in Section \ref{sec5}. Some feature of the other specific heats are discussed in appendix \ref{appB}.

  \section{NTG geometry } \label{sec2}
 As was pointed out in the introduction of this paper, it is important to construct an appropriate metric which explains the one-to-one correspondence between phase transitions and singularities of the scalar curvature. In Ref. \cite{HosseiniMansoori:2019jcs} we have proposed a new formalism of the thermodynamic geometry (NTG) which confirms this correspondence. 
  The NTG geometry is defined by
\begin{equation}\label{Ru1}
dl_{NTG}^2=\frac{1}{T}\left( \eta_i^{ j} \, \frac{\partial^2\Xi}{\partial X^j \partial X^l} \, d X^i d X^l \right)
\end{equation}
where $\eta_i^{ j}={\rm diag} (-1,1,...,1)$ and $\Xi$ is thermodynamic potential and $X^{i}$ can be intensive and extensive variables \cite{HosseiniMansoori:2019jcs}. It is interesting to note that the geometrothermodynamics (GTD) metric is conformally related to NTG metric such that this conformal transformation is singular at unphysical points were generated in GTD metric \cite{HosseiniMansoori:2019jcs}.

The NTG results extend further our knowledge of phase transition points.  For example, in two dimensional thermodynamic space, by selecting $\Xi=M(S,Q)$ and $X^{i}=(S,Q)$ in Eq. (\ref{Ru1}) 
 it is straightforward to see that curvature singularities correspond precisely to phase transitions of $C_{Q}$. In the same way, as one chooses thermodynamic potential by Legendre transformation like $\Xi=H(S,\Phi)=M-\Phi Q$, the  curvature singularity occurs exactly at the same location as the phase transition point of $C_{\Phi}$ \cite{HosseiniMansoori:2019jcs}. 
 The Legendre potentials in NTG formalism are obtained from the internal potential (or mass potential) by adding different combinations of
extensive and intensive variables. For instance, all thermodynamic potentials in two dimensional space are given by
\begin{eqnarray}
&&H(S,\Phi)=M(S,Q(S,\Phi))-\Phi Q(S,\Phi)\\
&& F(T,Q)=M(S(T,Q),Q)-TS(T,Q)\\
&& G(T,\Phi)=M(S(T,\Phi),Q(T,\Phi))-T S(T,\Phi)-\Phi Q(T,\Phi) 
\end{eqnarray}
where $H$, $F$, and $G$ are  enthalpy, free energy, and gibbs energy, respectively. In following, we shall demonstrate that NTG metrics coming from both free energy and entalpy potentials give the same result of the phase transition of $C_{\Phi}$.
 By choosing $\Xi=F(T,Q)$ with $X^{i}=(T,Q)$, NTG metric yields
 \begin{equation}\label{met1}
g^{NTG}_{F}=\frac{1}{T} \left( \begin{matrix}
-\Big(\frac{\partial^2 F}{\partial T^2}\Big)& 0\\
0 & \Big(\frac{\partial^2 F}{\partial Q^2}\Big)\\
\end{matrix}\right)=\frac{1}{T} \left( \begin{matrix}
\Big(\frac{\partial S}{\partial T}\Big)_{Q}& 0\\
0 & \Big(\frac{\partial \Phi}{\partial Q}\Big)_{T}\\
\end{matrix}\right)
 \end{equation}
 in which we have used the first law of thermodynamics for free energy, $dF=-SdT+\Phi dQ$. One can express the above metric elements in the new coordinate like $(S,\Phi)$ as 
 \begin{eqnarray}
 g_{TT}&=&\frac{1}{T} \Big(\frac{\partial S}{\partial T}\Big)_{Q}= \frac{1}{T} \frac{\{S,Q\}_{S,\Phi}}{\{T,Q\}_{S,\Phi}}\\
 g_{QQ}&=&\frac{1}{T} \Big(\frac{\partial \Phi}{\partial Q}\Big)_{T}= \frac{1}{T} \frac{\{\Phi,T\}_{S,\Phi}}{\{Q,T\}_{S,\Phi}}= -\frac{1}{T} \frac{\{\Phi,T\}_{S,\Phi}}{\{T,Q\}_{S,\Phi}}
\end{eqnarray}  
Appendix \ref{A1} is devoted to a brief introduction of the bracket notation. Moreover,  in the new coordinate $(S,\Phi)$, the metric elements must be changed by
\begin{equation}
\hat{g}=J^{T} g_{F}^{NTG} J
\end{equation}
where $J^{T}$ is the transpose of the Jacobian matrix $J$ which is defined as
\begin{equation}
J=\frac{\partial(T,Q)}{\partial (S,\Phi)}=\left( \begin{matrix}
\Big(\frac{\partial T}{\partial S}\Big)_{\Phi}& \Big(\frac{\partial T}{\partial \Phi}\Big)_{S}\\
\Big(\frac{\partial Q}{\partial S}\Big)_{\Phi} & \Big(\frac{\partial Q}{\partial \Phi}\Big)_{S}\\
\end{matrix}\right)
\end{equation}
Under varying coordinates and using Maxwell relation, $\Big(\frac{\partial T}{\partial \Phi}\Big)_{S}=-\Big(\frac{\partial Q}{\partial S}\Big)_{\Phi}$, the metric (\ref{met1}) takes the following form
\begin{equation}
\hat{g}=\frac{1}{T} \left( \begin{matrix}
\Big(\frac{\partial T}{\partial S}\Big)_{\Phi}& 0\\
0 & -\Big(\frac{\partial Q}{\partial \Phi}\Big)_{S}\\
\end{matrix}\right)=\frac{1}{T} \left( \begin{matrix}
\Big(\frac{\partial^2 H}{\partial S^2}\Big)& 0\\
0 & \Big(\frac{\partial^2 H}{\partial \Phi^2}\Big)\\
\end{matrix}\right)=-g^{NTG}_{H}
\end{equation}
In the last part, we have used the first law of thermodynamic for entalpy potential, i.e. $dH=TdS-Qd\Phi$. Clearly, their associated metrics are negative of each other, i.e.
\begin{equation}
g^{NTG}_{H}=-J^{T} g^{NTG}_{F} J \hspace{0.5cm} \text{or} \hspace{0.5cm} dl^{2}(H)=-dl^{2}(F)
\end{equation}
Therefore, the singularity of both $R^{F}$ and $R^{H}$ correspondences to the divergence of $C_{\Phi}$. 
It is worth mentioning that this result is true for conjugate potential pairs  $(\Xi, \overline{\Xi})$ which satisfy the following relation \cite{Liu:2010sz}.
\begin{equation}
\Xi+\overline{\Xi}=2 M-TS-\sum_{i}\Phi_{i} dQ_{i}
\end{equation}
Therefore, for conjugate pair $(M,G)$, one can prove that $g^{NTG}(M)=-N^{T} g^{NTG}_{G} N^{T}$ by using Jacobian matrix $N=\frac{\partial (T,\Phi)}{\partial (S,Q)}$.   Table \ref{tab:table1} presents the relation between curvature singularities and heat capacity phase transitions in three dimensional thermodynamic space with the first law, $dM=TdS+\Phi dQ+\Omega dJ$. 
\begin{table}[h!]
  \begin{center}
    \begin{tabular}{c|c|c} 
      \textbf{Thermodynamic potentials} & \textbf{ Jacobian matrix} & \textbf{Heat Capacities}\\
      $(\Xi,\overline{\Xi})$ & $g^{NTG}_{E}=-J^{T} g^{NTG}_{\overline{E}}J$ & $C$ \\
      \hline
      \hline  
       &  & \\
      $(M,M-TS-Q\Phi-\Omega J)$ & $\frac{\partial (T,\Phi,\Omega)}{\partial (S,Q,J)}$ & $C_{Q,J}$\\
     $(M-Q\Phi,M-TS-\Omega J)$ & $\frac{\partial (T,Q,\Omega)}{\partial (S,\Phi,J)}$ & $C_{\Phi,J}$\\
      $(M-\Omega J,M-TS-Q\Phi)$ & $\frac{\partial (T,\Phi,J)}{\partial (S,Q,\Omega)}$ & $C_{Q,\Omega}$\\
      $(M-Q\Phi-\Omega J,M-TS)$ & $\frac{\partial (T,Q,J)}{\partial (S,\Phi,\Omega)}$ & $C_{\Phi,\Omega}$\\
    \end{tabular}
     \caption{The relationship between curvature singularities and heat capacity divergences.\label{tab:table1} }
  \end{center}
\end{table}

Note that the coordinate transformation in our formalism can be interpreted as a special class of diffeomorphism invariant. In fact, diffeomorphism invariant of NTG formalism leads to introduce new thermodynamic potentials ($\bar \Xi$) which are related to the fundamental thermodynamic potentials ($\Xi$) by means of Legendre transformations as shown in Tab. \ref{tab:table1} \footnote{In Ref. \cite{Pineda:2017cgn}, it has been shown that there is a special diffeomorphism which transforms the partially and totally Legendre invariant metric of GTD into a Hessian metrics. One can get the same result in NTG formalism by inverting the Jacobin matrix of coordinate transformations in \cite{HosseiniMansoori:2019jcs}.}. 
 
  In the next section, we apply NTG geometry to finding the correspondence between curvature singularities and phase transitions for a 4D  charged Gauss Bonnet AdS black hole \cite{Fernandes:2020rpa} in the normal phase space where the AdS radius $l$ will be taken fixed. 
\section{Curvature singularities and phase transition signals in the normal phase space}\label{sec3}
 The Einstein-Maxwell-Gauss-Bonnet–Anti-de Sitter action in higher
dimensions $D$ can be written as \cite{Fernandes:2020rpa}
\begin{equation}
S=\frac{1}{16\pi} \int d^D x \sqrt{-g} \left[R+\frac{(D-1)(D-2)}{ l^2} +\alpha\mathcal{G}-F_{\mu \nu}F^{\mu \nu}\right]
\end{equation}
where $l$ is the AdS radius, $\alpha$ is the Gauss-Bonnet coefficient with dimension [length]$^{2}$, and $\mathcal{G}$ is the Gauss Bonnet invariant which is defined by 
\begin{equation}
\mathcal{G}=R^2-4 R_{\mu \nu}R^{\mu \nu}+R_{\mu \nu \rho \sigma}R^{\mu \nu \rho \sigma}
\end{equation}
 and the Maxwell field strength is defined by $F_{\mu \nu}=\partial_{\mu} A_{\nu}-\partial_{\nu} A_{\mu}$ where $A_{\mu}$ is the four vector potential. The spherically symmetric solution form its equations of motion after re-scaling the coupling constant by $\alpha/(D-4)$, in the limit $D \to 4$, takes the following form.
\begin{eqnarray}\label{solution}
ds^2&=&-f(r)dt^2+\frac{1}{f(r)}dr^2+r^2d\Omega ^2,\\
\nonumber f(r)&=&1+\frac{r^2}{2\alpha} \left(1-\sqrt{1+4 \alpha  \left(-\frac{1}{l^2}+\frac{2 M}{r^3}-\frac{Q^2}{r^4}\right)}\right)
\end{eqnarray}
in which $Q$ and $M$ are the charge and mass of the black hole \cite{Fernandes:2020rpa}. It is worth mentioning that the black-hole metric (\ref{solution}) has been previously  
 obtained in the semi-classical Einstein's equations with conformal anomaly \cite{Con1,Con2}, in gravity theory with quantum corrections \cite{Qon}, and also recently in the third order regularized Lovelock gravity \cite{Hor1, Hor2, Hor3, Hor4}. It should be also noted that this spherical black hole solution can be counted as a solution of the consistent theory proposed in Ref.\cite{Aoki:2020lig}. The explicit form for $M$ is obtained by using the condition $f(r_+)=0$ as
\begin{equation}\label{mass1}
M=\frac{r_{+}^3}{2 l^2}+\frac{Q^2}{2 r_{+}}+\frac{\alpha }{2 r_{+}}+\frac{r_{+}}{2}.
\end{equation}
Note that the first law of thermodynamic, $dM=TdS+....$ is always satisfied for this black hole, here ... stands for some work terms. By integrating the first law, the entropy for this black hole can be obtained as
\begin{equation}\label{entropy}
S=\int T^{-1} \Big(\frac{\partial M}{\partial r_{+}}\Big)_{Q,\alpha} dr_{+}=\pi  r_+^2+4 \alpha \pi \ln (r_+)+S_{0}
\end{equation}
where $S_{0}$ is an integration constant, which we can not fix it due to the existence of the logarithmic term. The black hole entropy can also be written in the term of horizon area $A=4 \pi r_{+}^2$ as
\begin{equation}\label{Entropy1}
S=\frac{A}{4}+2 \pi \alpha \ln \Big(\frac{A}{A_{0}}\Big)
\end{equation}
where $A_{0}$ is a constant with dimension of area. Clearly, there exists a logarithmic correction to the well-known Bekenstein-Hawking area entropy in comparison with RN-AdS case.  It is worthwhile noting that such a logarithmic term also appears in the entropy formula of some quantum theories of gravity such as loop quantum gravity and string theory \cite{Con1,Con2,Qon}.

Using the first law of thermodynamics, $dM=TdS+\Phi dQ$, Hawking temperature ($T$),  electric potential  ($\Phi$), and specific heat capacity at constant electric charge ($C_{Q}$) are given by

 \begin{equation} \label{mm2}
 T={(\frac{\partial M}{\partial S})}_{Q}=\frac{\{M,Q\}_{r_{+},Q}}{\{S,Q\}_{r_{+},Q}}=\frac{3 r_{+}^4-l^2 (Q^2-r_{+}^2+\alpha)}{4 l^2 \pi r_{+}(r_{+}^2+2 \alpha)}
 \end{equation}
 \begin{equation}\label{Phi1}
 \Phi ={(\frac{\partial M}{\partial Q})}_{S}=\frac{\{M,S\}_{r_{+},Q}}{\{Q,S\}_{r_{+},Q}}=\frac{Q}{r_{+}}
 \end{equation}
 \begin{equation}  \label{ee6}
 {{C}_{Q}}=T{{(\frac{\partial S}{\partial T})}_{Q}}=T\frac{\{S,Q\}_{r_{+},Q}}{\{T,Q\}_{r_{+},Q}}=\frac{2 \pi (r_{+}^2+2 \alpha)^2 (3 r_{+}^4-l^2(Q^2-r_{+}^2+\alpha))}{3 (r_{+}^6+6 r_{+}^4 \alpha)+l^2 (-r_{+}^4+5 r_{+}^2 \alpha+2 \alpha^2+Q^2 (3 r_{+}^2+2 \alpha))}
\end{equation}
Let us know how the NTG method can reveal the critical behavior of heat capacity $C_{Q}$. To do this, one needs to plug the thermodynamic potential $\Xi=M(S,Q)$ with $X^{i}=(S,Q)$ into Eq. (\ref{Ru1}), i.e.,
  \begin{equation}
  (dl^{NTG})^{2}=\frac{1}{T}\left(-\frac{\partial^2 M}{\partial S^2}dS^2+\frac{\partial^2 M}{\partial Q^2}dQ^2\right)\
  \end{equation}
  thus the denominator of $R^{NTG}$ reads
 \begin{eqnarray} \label{ee5}
\nonumber D(R^{NTG}) &=&\pi\Big((r_{+}^2+2 \alpha)^2 (3 r_{+}^4- l^2(Q^2-r_{+}^2-\alpha))\Big) \\
&\times& \Big(3 (r_{+}^6+6 r_{+}^4 \alpha)+l^2 (-r_{+}^4+5 r_{+}^2 \alpha+2 \alpha^2+Q^2 (3 r_{+}^2+2 \alpha)\Big)^2
 \end{eqnarray}
 \begin{figure}[tbp]
 \includegraphics[scale=0.4]{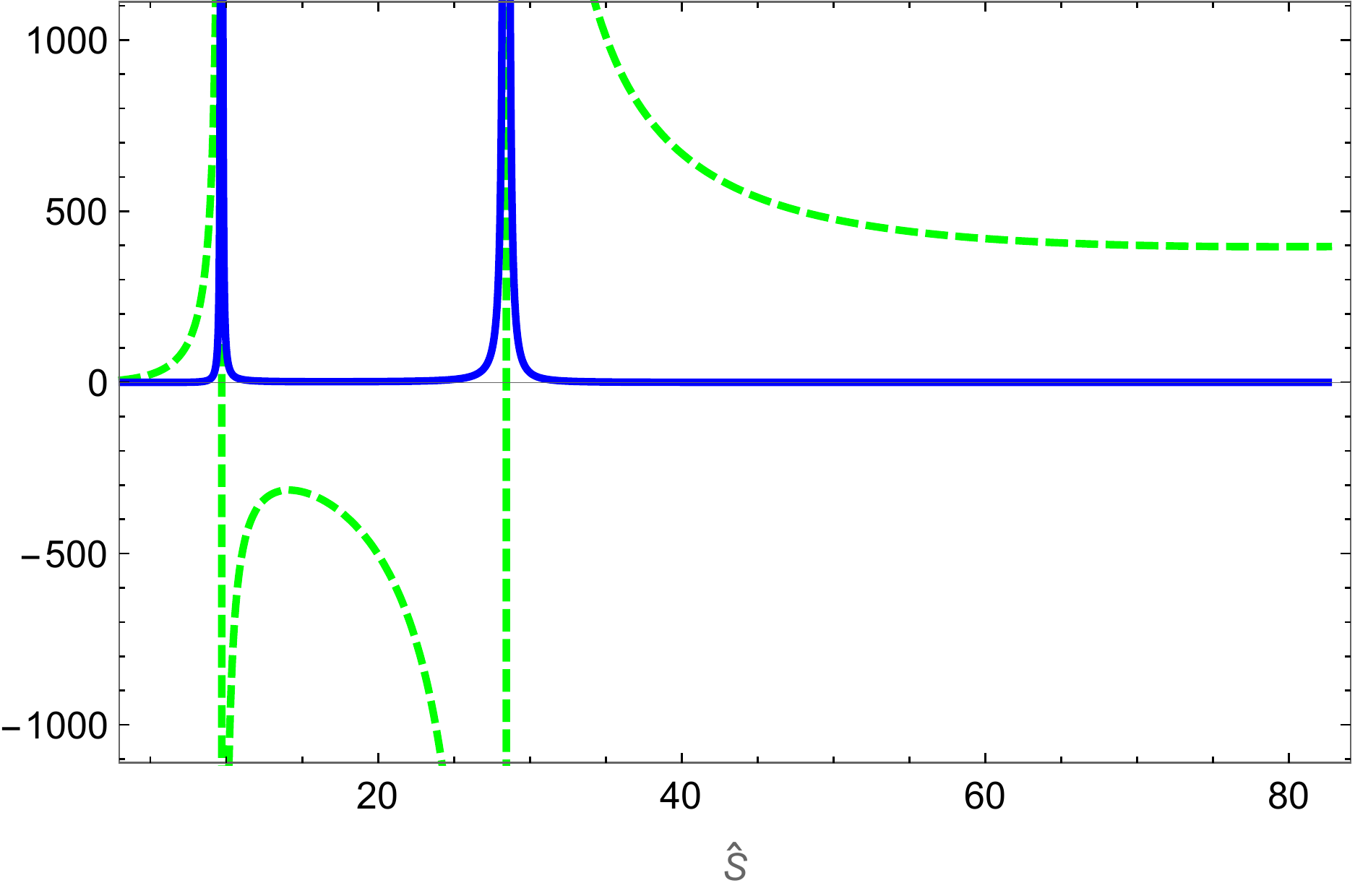}
  \includegraphics[scale=0.4]{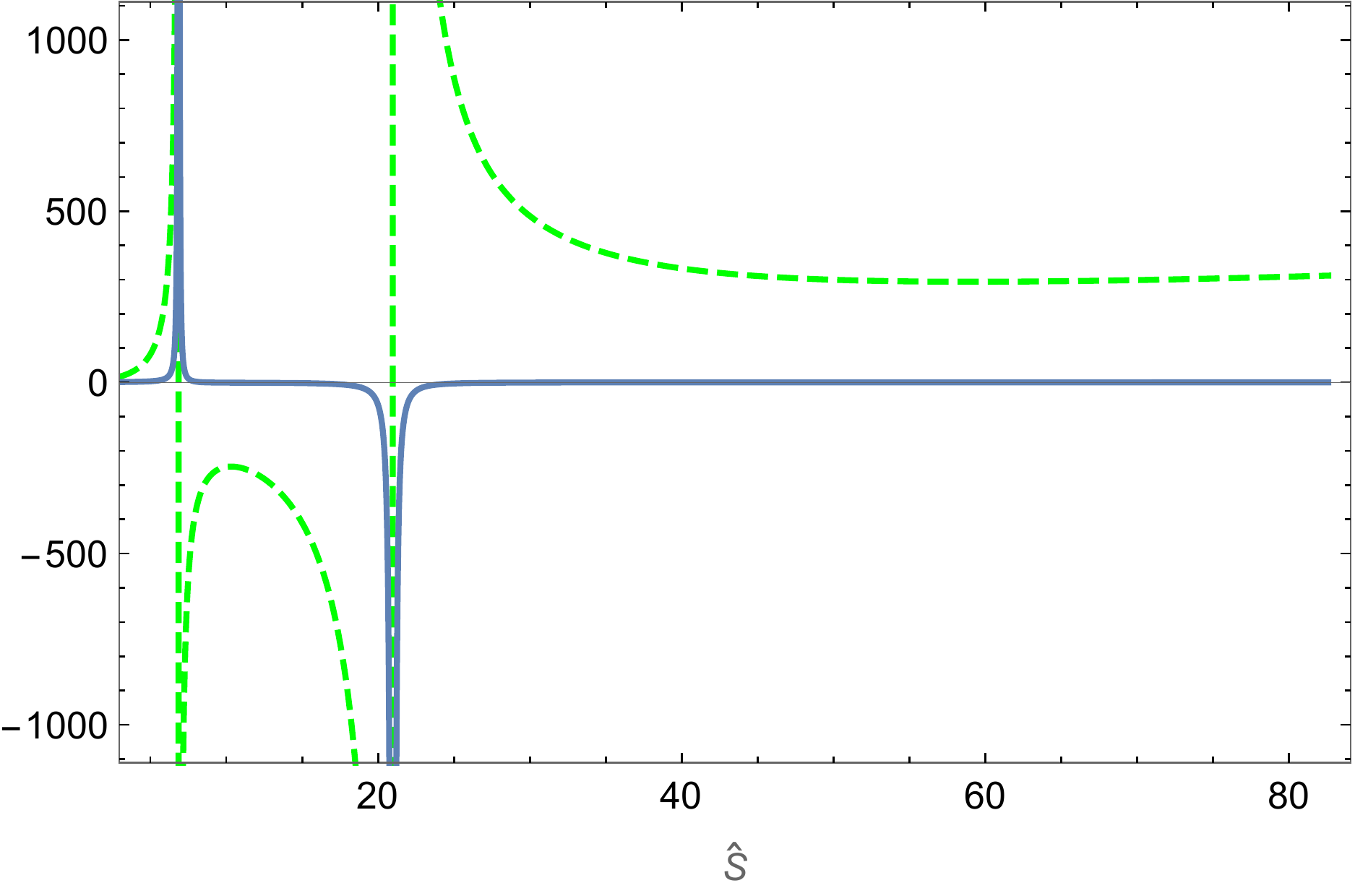}
 \caption{Left: The diagram of the phase transition of $C_{Q}$ (dashed green curve) and the scalar curvature  $R^{NTG}(\hat{S},Q)$ (solid blue curve) with respect to entropy, $\hat{S}=S-S_{0}$, with $Q=0.5$ for a charged GB-AdS black hole. Right: The diagram of the phase transition of $C_{\Phi}$ (dashed green curve) and the scalar curvature  $R^{NTG}(\hat{S},\Phi)$ (solid blue curve) with respect to entropy, $\hat{S}$, with $\Phi=0.5$ for a charged GB-AdS black hole. In both diagrams, we have considered
   $l=6$, and $\alpha=0.2$. \label{fig1}   }
 \end{figure}
It is obvious that the leading term in the denominator is zero only at the extremal limit ($T = 0$) which is forbidden by the third law of thermodynamics, while the roots of the second part give us all phase transition points of $C_{Q}$. This result has been illustrated in the left hand side of Fig. (\ref{fig1}). It should be noted that positive regions of heat capacity diagram correspond to a stable system whereas negative regions indicate  instability of a system. As a consequence of the NTG method, curvature singularities occur exactly at phase transitions with no other additional roots.
 Furthermore, by using Eq. (\ref{Phi1}) for $Q$, one can define the heat capacity at fixed electric potential as
\begin{eqnarray}
C_{\Phi}  =T\frac{\{S,\Phi\}_{r_{+},\Phi}}{\{T,\Phi\}_{r_{+},\Phi}}=\frac{2 \pi (r_{+}^2+2 \alpha)^2(3 r_{+}^4-l^2(\alpha+r_{+}^2(-1+\Phi^2)))}{3 (r_{+}^6+6 r_{+}^4 \alpha)+l^2 (2 \alpha^2+r_{+}^2 \alpha (5-2 \Phi^2)+r_{+}^4 (-1+\Phi^2))}
\end{eqnarray}
Let us now construct NTG metric in this case. Starting with NTG metric (\ref{Ru1}), and considering $\Xi=H(S,\Phi)=M(S,Q(S,\Phi))-Q(S,\Phi) \Phi$ and $X^{i}=(S,\Phi)$, we arrive at
\bea
dl_{NTG}^2= \frac{1}{T} \Big(-\frac{\partial^2 H}{\partial S^2}dS^2+\frac{\partial^2 H}{\partial \Phi^2}d\Phi^2\Big)
\eea
Thus the denominator of the scalar curvature is
\begin{eqnarray}
D(R^{NTG}) &=& \pi (r_{+}^2+2 \alpha)^2 (-3 r_{+}^4+l^2 (\alpha+r_{+}^2(-1+\Phi^2))) \\
\nonumber &\times & \Big(3 (r_{+}^6+6 r_{+}^4 \alpha)+l^2 (2 \alpha^2+r_{+}^2 \alpha (5-2 \Phi^2)+r_{+}^4 (-1+\Phi^2))\Big)^2
\end{eqnarray}
Remarkably, the curvature singularities give us the phase transition points of $C_{\Phi}$ (See the right hand side of Fig. (\ref{fig1})). In summary, NTG geometry can provide us with a powerful tool to achieve a one-to-one correspondence between singularities and phase transitions.

\begin{figure}[tbp]
 \includegraphics[scale=0.4]{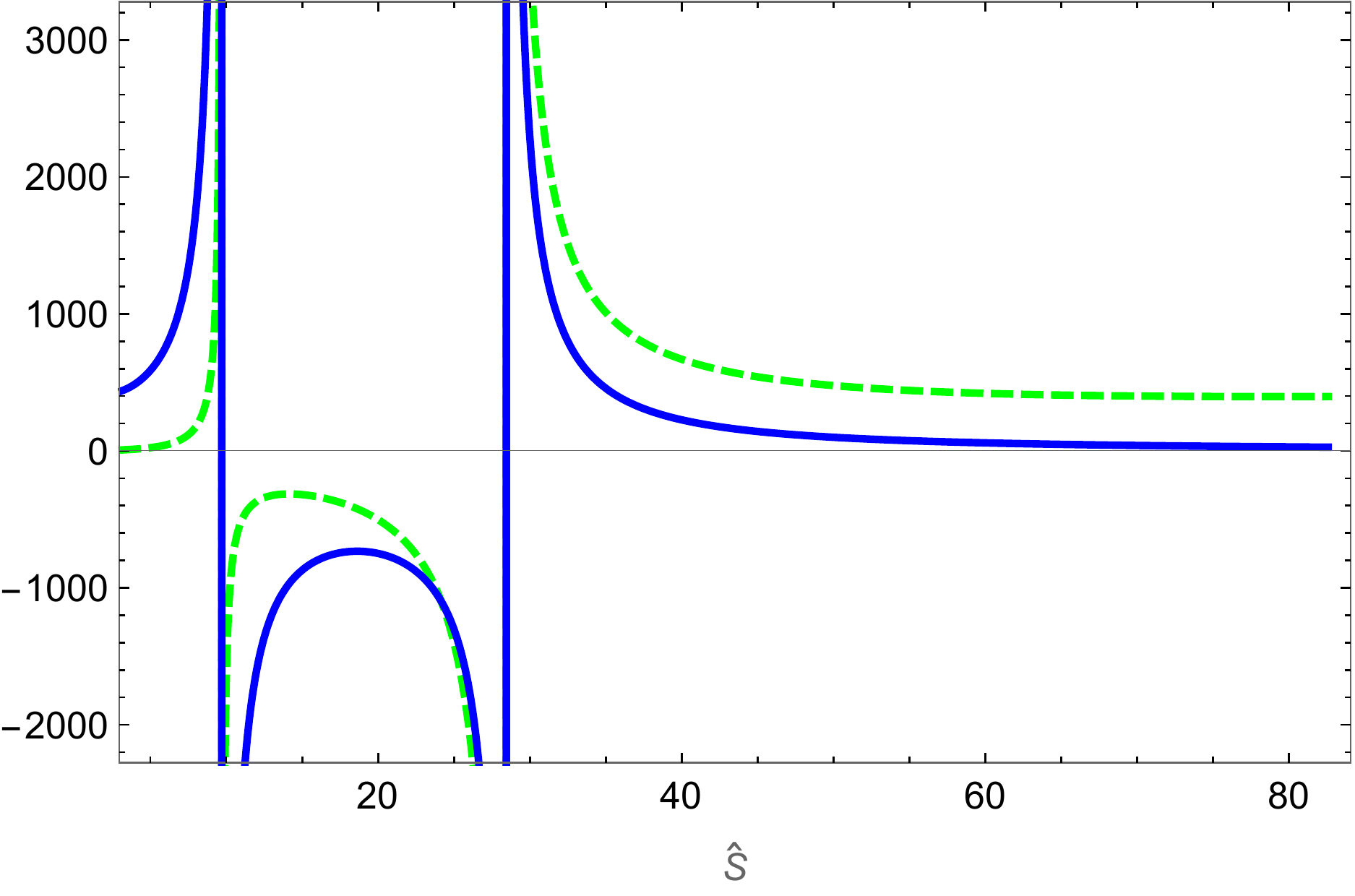}
 \includegraphics[scale=0.4]{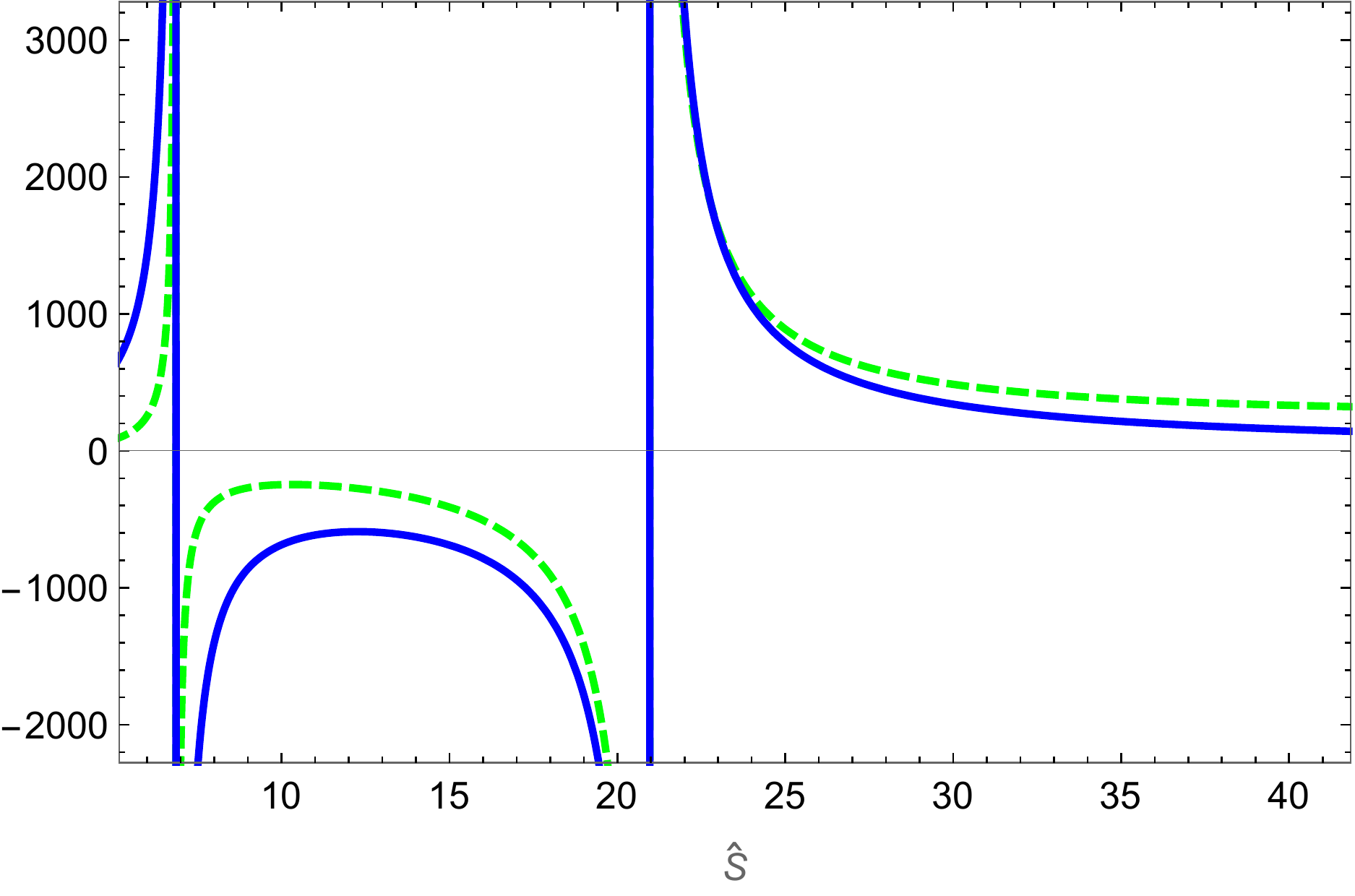}
 \caption{Left: The diagram of the phase transition of $C_{Q}$ (dashed green curve) and the extrinsic curvature  $K^{NTG}(\hat{S},Q)$ ($\times 10^3$) (solid blue curve) with respect to entropy, $\hat{S}=S-S_{0}$, with $Q=0.5$ for a charged GB-AdS black hole. Right: The diagram of the phase transition of $C_{\Phi}$ (dashed green curve) and the extrinsic curvature  $K^{NTG}(\hat{S},\Phi)$ ($\times 10^3$) (solid blue curve) with respect to entropy, $\hat{S}$, with $\Phi=0.5$ for a charged GB-AdS black hole. In both diagrams, we have considered
  $l=6$, and $\alpha=0.2$. \label{fig2}  }
 \end{figure}

Although NTG curvature determines where phase transition points occur, it fails to explain thermal stability of a thermodynamic system. In Ref. \cite{Mansoori:2016jer} we have shown that the extrinsic curvature of a certain kind of hypersurface immersed in thermodynamic space contains unexpected information about stability of a thermodynamic system. 
Strictly speaking, the extrinsic curvature of such a hypersurface not only is singular at phase transition points, but also has the same sign as the heat capacity around the phase transition points.
Let us briefly review the basic concept of the extrinsic curvature in the thermodynamic manifold. For a D-dimensional thermodynamic manifold $\mathcal{M}$ with coordinate $X^{i}$, a special hypersurface $\Sigma$ embedded in $\mathcal{M}$ is defined by the surface equation $\mathcal{P}(X^{i})=0$ and the orthogonal normal vector \cite{book},
\begin{equation}
n_{\mu}=\frac{\partial_{\mu} \mathcal{P}}{\sqrt{|\partial_{\mu} \mathcal{P} \partial^{\mu} \mathcal{P}|}}
\end{equation}
Therefore, the extrinsic curvature tensor on this hypersurface is defined as \cite{Mansoori:2016jer}
\begin{equation}\label{Ex1}
K=\nabla_{\mu} n^{\mu}=\frac{1}{\sqrt{g}} \partial_{\mu} \Big(\sqrt{g} n^{\mu}\Big)
\end{equation} 
The most striking result is that extrinsic curvature indicates perfectly stability/instability of thermal phase transition \cite{Mansoori:2016jer}. In order to see the behavior of $C_{Q}$ around phase transitions, we must restrict ourselves to living on a constant $Q$ hypersurface with the normal vector $n_{Q}=\frac{1}{\sqrt{{g^{NTG}}^{QQ}}}$. From Eq. (\ref{Ex1}), we have 
\begin{equation}
K^{NTG}=\frac{\sqrt{r_{+}^2+2\alpha}\Big(l Q r_{+}^2(l^2+6 r_{+}^2)\Big)\Big(|\pi (3 r_{+}^4-l^2 (Q^2-r_{+}^2+\alpha))|\Big)^{-\frac{1}{2}}}{3 (r_{+}^6+6 r_{+}^4\alpha)+l^2 (-r_{+}^4+2 \alpha (Q^2+\alpha)+r_{+}^2(3 Q^2+5\alpha))}
\end{equation}
Transparently, the denominator term indicates the phase transition points, whereas the second term in numerator is only zero at $T = 0$. The left hand side of Fig. \ref{fig2} pinpoints exactly the extrinsic curvature has the same sign as heat capacity does. Nonetheless this result was not expected for the scalar curvature as shown in the left hand side of Fig. \ref{fig1}. 
This finding can be predicted for $C_{\Phi}$ when we consider a constant $\Phi$ hypersurface with unit normal vector $n_{\Phi}=\frac{1}{\sqrt{{g^{NTG}}^{\Phi \Phi}}}$. Utilizing Eq. (\ref{Ex1}), the extrinsic curvature is obtained to be
\begin{equation}
K^{NTG}=\frac{\Big(l r_{+} \sqrt{r_{+}^2+2 \alpha}(3 r_{+}^4+l^2 \alpha) \Phi\Big) \Big(|\pi (3 r_{+}^4-l^2 (\alpha+r_{+}^2(\Phi^2-1)))|\Big)^{-\frac{1}{2}}}{ 3 (r_{+}^6+6 r_{+}^4 \alpha)+l^2 (2 \alpha^2+r_{+}^2 \alpha (5-2 \Phi^2)+r_{+}^4(\Phi^2-1))}
\end{equation}
The right hand side of Fig. (\ref{fig2}) illustrates that extrinsic curvature diverges at 
 phase transition points and exhibits a similar behavior around such points as the heat capacity does.

 \section{Curvature singularities and phase transition signals in the extend phase space }\label{sec4}
 In this section, we examine NTG geometry for one 4D charged GB-AdS black hole in the extended phase space where the cosmological constant (or the AdS radius) treats as thermodynamic pressure, i.e. $P=\frac{3}{8 \pi l^2}$ \cite{Kubiznak:2012wp}. Expressing the AdS radius $l$ in terms of the pressure $P$, the first law for the black hole is defined as
\begin{equation}\label{smarr}
dM=TdS+VdP+\Phi dQ+ \mathcal{A}d\alpha 
\end{equation}
where $V$ and $\mathcal{A}$ are thermodynamic
quantities conjugating to pressure $P$ and Gauss Bonnet coupling coefficient $\alpha$, respectively. According to Eq. (\ref{smarr}), we find the black hole mass $M$ should be treated as enthalpy, i.e., $M\equiv H $ rather than internal energy $E$ \cite{Kubiznak:2012wp}. Taking advantage of Eqs. (\ref{mass1}) and (\ref{smarr}), thermodynamic volume is given by
\begin{equation}
V=\left( \frac{\partial M}{\partial P}\right) _{S,Q,\alpha}=\frac{4}{3} \pi  r_+^3=\frac{\pi}{6} v^3
\end{equation}
where $v = 2r_{+}$ denotes the specific volume with dimension of length. It is trivial to verify that  
\begin{equation}
F=E-TS=M-PV-TS=\frac{4 Q^2+v^2-\pi T v^3+4 \alpha-16\alpha \pi T v \ln(\frac{v}{v_{0}}) }{4v}
\end{equation}
 where $F$ stands for free energy \footnote{Note that the black hole entropy (\ref{Entropy1}) can be expressed in terms of special volume $v$ as
 \begin{equation}
 	S=\frac{\pi v^2}{4}+2 \pi \alpha \ln \Big(\frac{ \pi v^2}{A_{0}}\Big)
 \end{equation}	
 	in which we choose $A_{0}=\pi v_{0}^2$ with $v_{0}$ constant with dimension of length, which is needed to make the logarithm argument dimensionless.  }. On combing this result with the differential form for free energy, $dF=-SdT-PdV+\Phi dQ+\mathcal{A} d\alpha$, we deduce 
\begin{eqnarray}
S&=&-\Big(\frac{\partial F}{\partial T}\Big)_{V,Q,\alpha}=\frac{\pi}{4} \Big(v^2+16 \alpha \ln\Big(\frac{v}{v_{0}}\Big)\Big)\\
\label{P1} P&=&-\Big(\frac{\partial F}{\partial V}\Big)_{T,Q,\alpha}=-\frac{2}{\pi v^2}\Big(\frac{\partial F}{\partial v}\Big)_{T,Q,\alpha}=\frac{2 Q^2}{\pi v^4}-\frac{1}{2 \pi v^2}+\frac{T}{v}+\frac{2 \alpha}{\pi v^4}+\frac{8 T \alpha}{v^3}\\
\Phi&=&\Big(\frac{\partial F}{\partial Q}\Big)_{T,V,\alpha}=\frac{2 Q}{v}\\
\mathcal{A}&=&\Big(\frac{\partial F}{\partial \alpha}\Big)_{T,V,Q}=\frac{1}{v}-4 \pi T \ln\Big(\frac{v}{v_{0}}\Big)
\end{eqnarray} 
Notice that the form of Eq. (\ref{P1}) is reminiscent of the state equation for the Van der
Waals gas. In this regard, in Ref. \cite{Hegde:2020xlv} the authors have shown that there exists a small-large
black hole phase transition of VdW type for a 4D GB black hole case via isotherms in $P$- $V$ diagram. Moreover, the critical point can be obtained by solving  $(\partial_{v} P)_{T}=(\partial_{v,v}P)_{T}=0$, which gives \cite{Hegde:2020xlv}
\begin{eqnarray}\label{criticalpoints}
T_c&=&\frac{\left(8 \alpha +3 Q^2-\sqrt{48 \alpha ^2+9 Q^4+48 \alpha  Q^2}\right) \sqrt{6 \alpha +3 Q^2+\sqrt{48 \alpha ^2+9 Q^4+48 \alpha  Q^2}}}{48 \pi  \alpha ^2}\\
v_c&=& 2 \left(6 \alpha +3 Q^2+\sqrt{48 \alpha ^2+9 Q^4+48 \alpha  Q^2}\right)^{1/2}.
\end{eqnarray}
  The signal of a phase transition typically arises when a  specific heat capacity changes its sign, which indicates whether a system is stable or not. In the other words, a positive heat capacity implies stability of a thermal system whereas a negative heat capacity shows instability of such a system under imposing small perturbations.
Making use of above equations, the specific heat at constant pressure, electric charge, and GB coupling is given by
\begin{equation}
C_{P,Q,\alpha}=T\Big(\frac{\partial S}{\partial T}\Big)_{P,Q,\alpha}=T \frac{\{S,P,Q,\alpha\}_{T,v,Q,\alpha}}{\{T,P,Q,\alpha\}_{T,v,Q,\alpha}}=\frac{\pi^2T v(v^2+8 \alpha)^2}{2\Big(8Q^2+v^2(\pi T v-1)+8\alpha (1+3 \pi T v)\Big)}
\end{equation}
where stability requires $C_{P,Q,\alpha}>0$. It may easily verified that the specific heat $C_{P,Q,\alpha}$
becomes singular exactly at the critical point given by Eq. (\ref{criticalpoints}). 

Having enthalpy potential in our hand, we are able to implement NTG geometry to analysis phase transition behavior of $C_{P,Q,\alpha}$. By substituting  thermodynamic potential $\Xi=H=M=E+PV$ \footnote{The conjugate potential, $\overline{\Xi}=E-TS-\Phi Q-\alpha \mathcal{A}$ give us the same result.} with $X^{i}=(S,P,Q,\alpha)$ into Eq. (\ref{Ru1}), we have
\begin{equation}
g^{NTG}_{H}=\frac{1}{T}\left(\begin{matrix}
-H_{SS}& 0& 0 & 0\\
0& H_{PP}& H_{PQ} & H_{P \alpha}\\
0& H_{QP}& H_{QQ} & H_{Q \alpha}\\
0& H_{\alpha P}& H_{\alpha Q} & H_{\alpha \alpha}\\
\end{matrix}\right)
\end{equation}
Since all thermodynamic parameters are written as a function of $(T,v,Q,\alpha)$, it is convenient to recast metric elements from the coordinate $X^{i}=(S,P,Q,\alpha)$ to the favorite coordinate $(T,v,Q,\alpha)$. To do this, we first need to redefine metric elements as follows,
\begin{eqnarray}
\nonumber H_{SS}&=&\Big(\frac{\partial T}{\partial S}\Big)_{P,Q,\alpha}=\frac{\{T,P,Q,\alpha\}_{T,v,Q,\alpha}}{\{S,P,Q,\alpha\}_{T,v,Q,\alpha}}=-\frac{16 Q^2-2 v^2+2 \pi T v^3+16 \alpha+48 \pi T v \alpha}{\pi^2  v (v^2+8 \alpha)^2}\\
\nonumber H_{PP}&=&\Big(\frac{\partial V}{\partial P}\Big)_{S,Q,\alpha}=\frac{\{V,S,Q,\alpha\}_{T,v,Q,\alpha}}{\{P,S,Q,\alpha\}_{T,v,Q,\alpha}}=0\\
\nonumber H_{QQ}&=&\Big(\frac{\partial \Phi}{\partial Q}\Big)_{S,P,\alpha}=\frac{\{\Phi,S,P,\alpha\}_{T,v,Q,\alpha}}{\{Q,S,P,\alpha\}_{T,v,Q,\alpha}}=\frac{2}{ v}\\
 H_{\alpha \alpha}&=&\Big(\frac{\partial \mathcal{A}}{\partial \alpha}\Big)_{S,P,Q}=\frac{\{\mathcal{A},S,P,Q\}_{T,v,Q,\alpha}}{\{\alpha,S,P,Q\}_{T,v,Q,\alpha}}\\
\nonumber &=& 16 \ln \Big(\frac{v}{v_{0}}\Big) \frac{(1+4 \pi T v)(v^2+8 \alpha)+2(8 Q^2+v^2 (-1+\pi T v)+8 (1+3 \pi T v)) \ln \Big(\frac{v}{v_{0}}\Big)}{ v (v^2+8 \alpha)^2}\\
\nonumber H_{PQ}&=&H_{QP}=\Big(\frac{\partial V}{\partial Q}\Big)_{S,P,\alpha}=\frac{\{V,S,P,\alpha\}_{T,v,Q,\alpha}}{\{Q,S,P,\alpha\}_{T,v,Q,\alpha}}=0\\
\nonumber H_{P\alpha}&=&H_{\alpha P}=\Big(\frac{\partial \mathcal{A}}{\partial P}\Big)_{S,Q,\alpha}=\frac{\{\mathcal{A},S,Q,\alpha\}_{T,v,Q,\alpha}}{\{P,S,Q,\alpha\}_{T,v,Q,\alpha}}=-\frac{4 \pi v^3 \ln \Big(\frac{v}{v_{0}}\Big)}{v^2+8 \alpha}\\
\nonumber H_{Q\alpha}&=&H_{\alpha Q}=\Big(\frac{\partial \mathcal{A}}{\partial Q}\Big)_{S,P,\alpha}=\frac{\{\mathcal{A},S,P,\alpha\}_{T,v,Q,\alpha}}{\{Q,S,P,\alpha\}_{T,v,Q,\alpha}}=\frac{16 Q \ln \Big(\frac{v}{v_{0}}\Big)}{v(v^2+8 \alpha)}
\end{eqnarray}
 then transferring from coordinate $(S,P,Q,\alpha)$ to $(T,v,Q,\alpha)$ by using the below Jacobian matrix,
 \begin{equation}
 J=\frac{\partial (S,P,Q,\alpha)}{\partial (T,v,Q,\alpha)}=\left(\begin{matrix}
 0& \pi \Big(\frac{v}{2}+\frac{4 \alpha}{v}\Big)&0& 4 \pi \ln\Big(\frac{v}{v_{0}}\Big)\\
 \frac{v^2+8 \alpha}{v^3} & \frac{-8 Q^2+v^2-\pi T v^3-8 \alpha (1+3 \pi T v)}{\pi v^5} &  \frac{4 Q}{\pi v^4} & \frac{2+8 \pi T v}{\pi v^4}\\
 0 & 0 & 1 & 0\\
 0 & 0 & 0 & 1\\
 \end{matrix}\right)
 \end{equation}
finally the metric elements convert to 
 \begin{equation}\label{metB}
 \hat{g}=J^{T} g^{NTG}_{H} J=\left( \begin{matrix}
 0 & 0 & 0 & -\frac{4 \pi}{T} \ln \Big(\frac{v}{v_{0}}\Big)\\
 0& -\frac{8 Q^2+v^2 (\pi T v-1)+8\alpha (1+3 \pi T v)}{2 T v^3}&0 &0 \\
 0&0&\frac{2}{T v}&0\\
- \frac{4 \pi }{T} \ln \Big(\frac{v}{v_{0}}\Big)&0&0&0\\
 \end{matrix}\right)
 \end{equation}
This allows the denominator of the scalar
curvature to be
\begin{equation}\label{R31}
D(R^{NTG})=\pi \Big(8Q^2+v^2(\pi T v-1)+8\alpha (1+3 \pi T v)\Big)^2 \ln\Big(\frac{v}{v_{0}}\Big)^2
\end{equation}
Clearly, the first parenthesis presents phase transitions of $C_{P,Q,\alpha}$, whereas the logarithmic term gives us an extra singularity at $v=v_{0}$. The foremost cause of this discrepancy in the desired correspondence is a result of a notable quantum effect arose in 4D GB black holes. More details on this will be given below. 

\begin{figure}[tbp]
\includegraphics[scale=0.5]{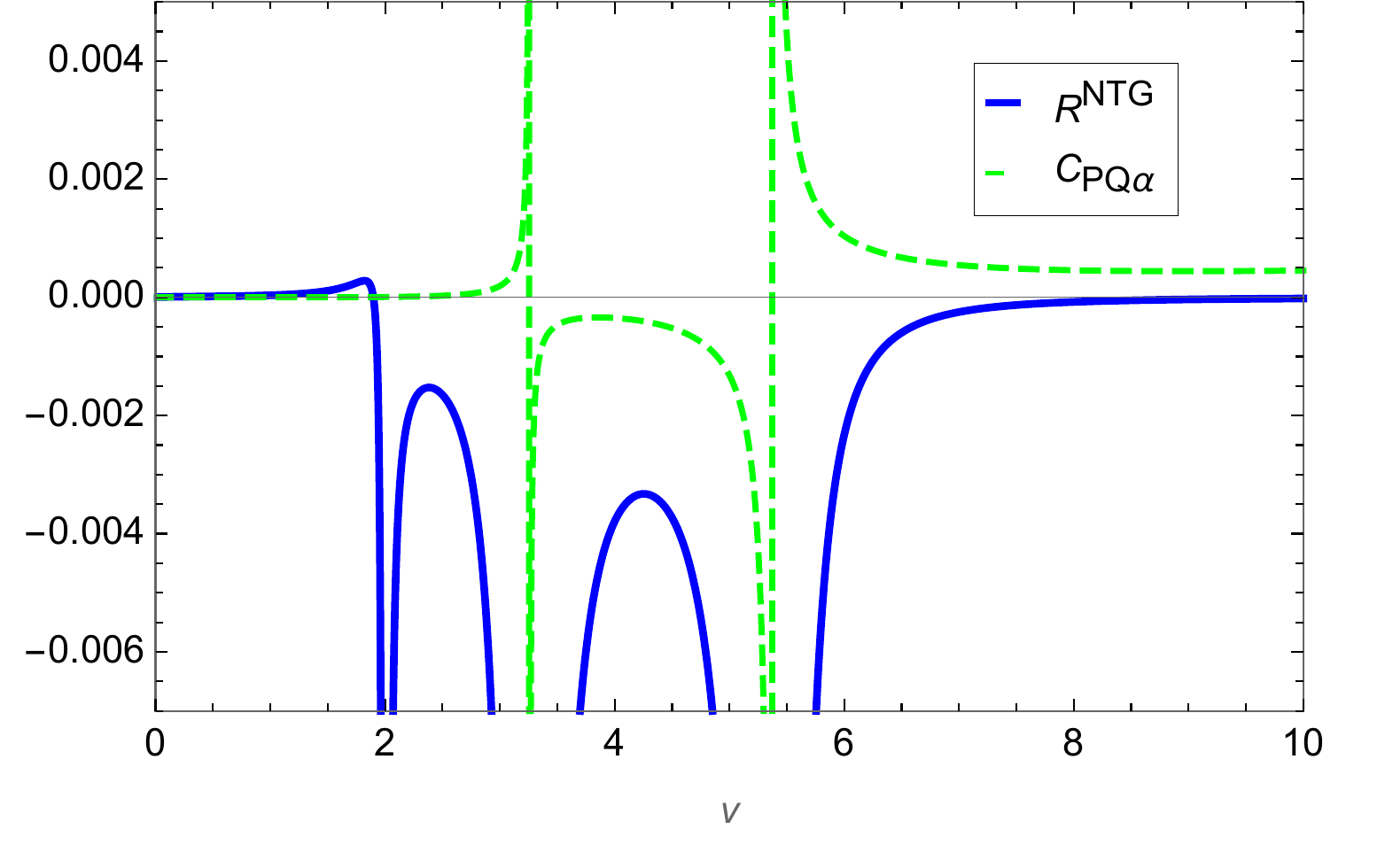}
\includegraphics[scale=0.5]{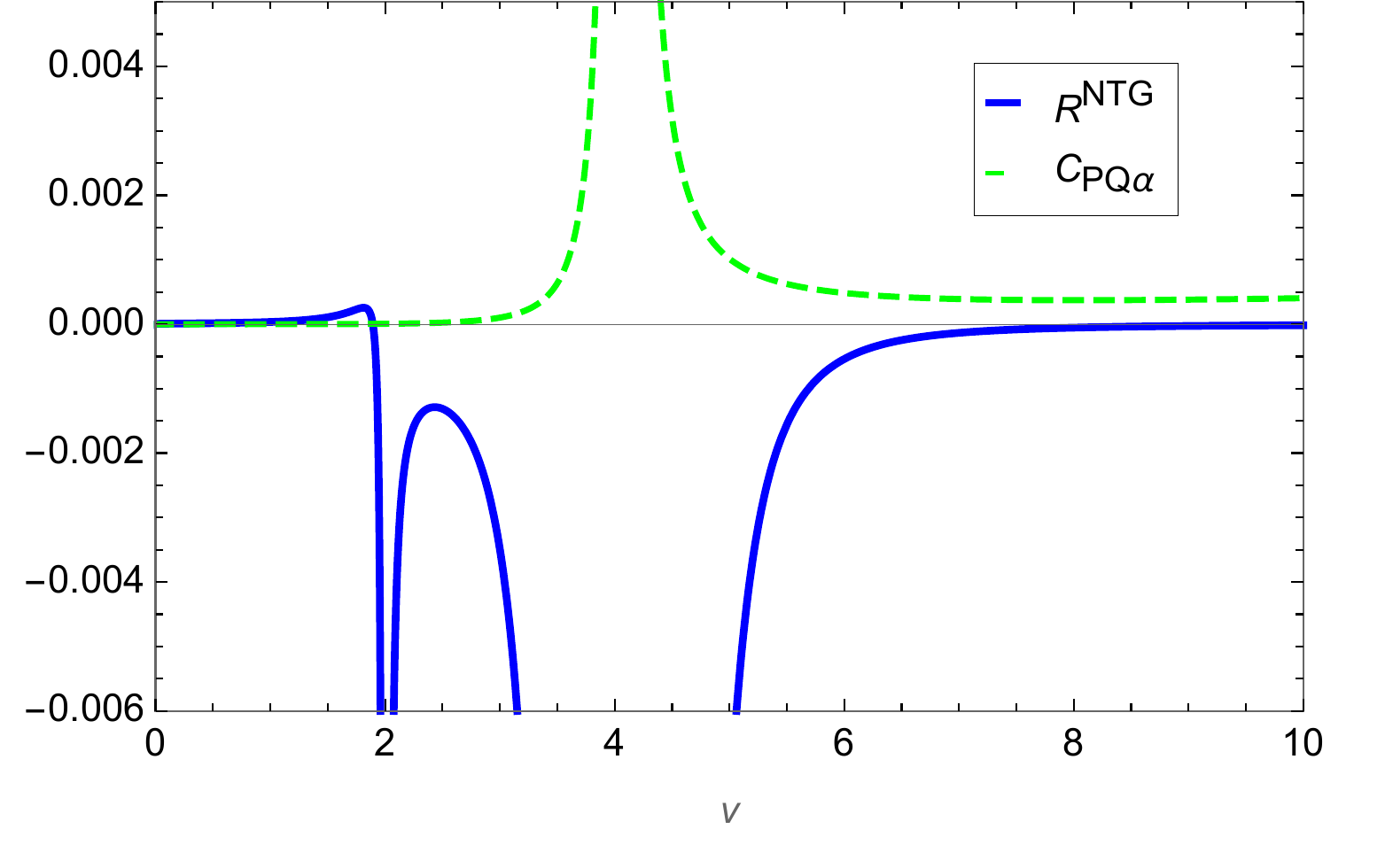}
\begin{center}
\includegraphics[scale=0.5]{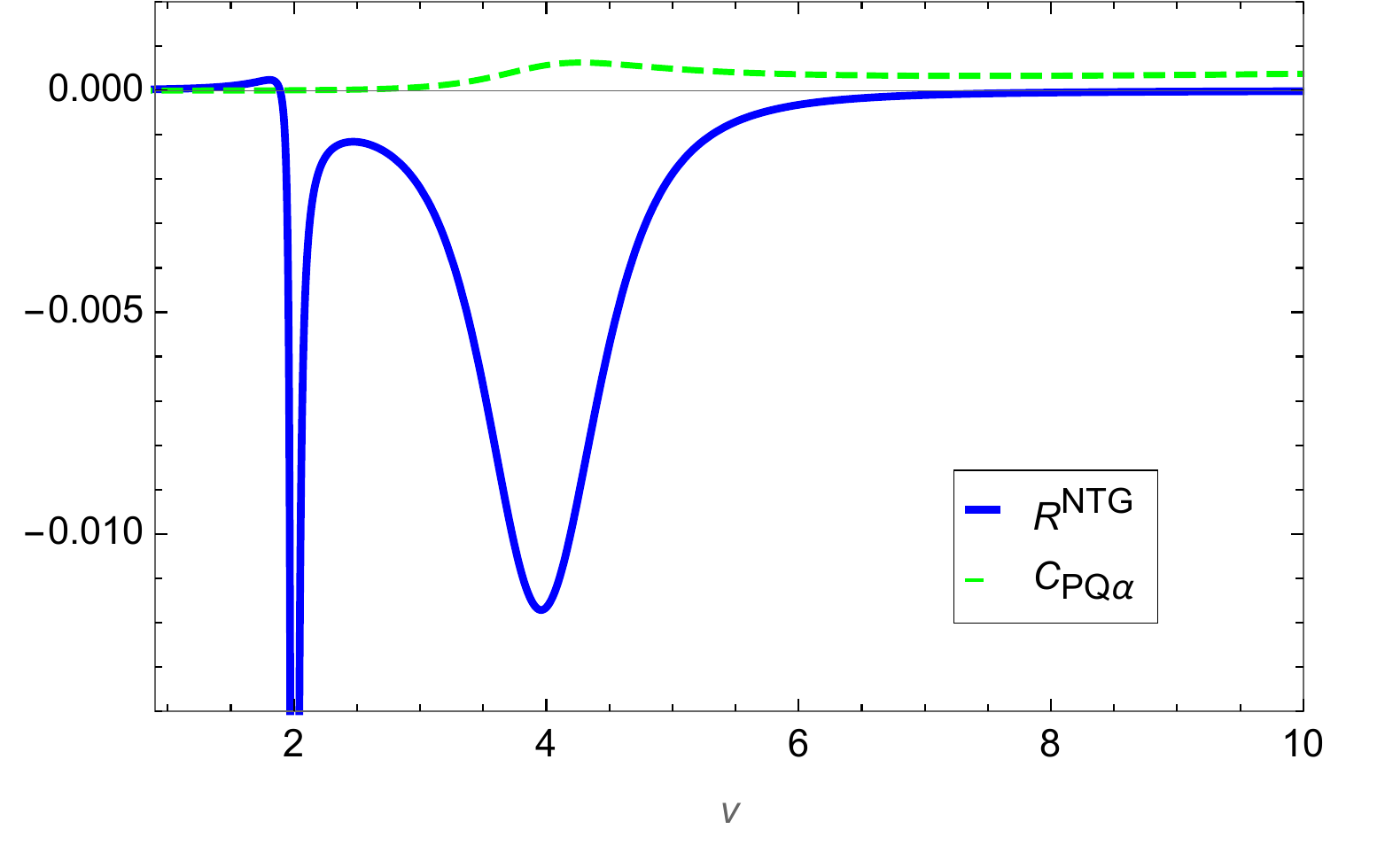}
\end{center}
\caption{The diagram of the specific heat ($\times 10^{6}$) and scalar curvature ($\times 10^{4}$) versus the specific volume $v$ for a 4D GB- AdS black hole. Form left to right we consider $T=\{0.0444,0.0474,0.0494\}$ where $T_c=0.0474$ for constant values of the electric charge $Q=0.5$
and Gauss-Bonnet coupling $\alpha=0.2$. Here, we have taken $v_{0}=2$. \label{RCPQa} }
\end{figure}

We have depicted
the scalar
curvature and heat capacity $C_{P,Q,\alpha}$ with respect to specific volume in Fig. \ref{RCPQa}.
We observe from it that for $T<T_{c}$ there exists two divergent points for $C_{P,Q,\alpha}$. The stable phases with positive specific heat happen in the lower and higher volume regions, while the intermediate phase with negative $C_{P,Q,\alpha}$ value is unstable phase. As temperature is smaller than its critical value there are three possible phases, i.e., the small black hole (SBH), intermediate black hole (IBH) and large black hole (LBH). By increasing temperature to $T=T_{c}$, these two divergent points get closer and coincide at $v=v_{c}=4.079$ to form a single divergence where the unstable region disappears. For $T>T_c$, heat capacity is always positive and the divergent point vanishes. It means that the black hole is stable and there is no phase transition. Moreover, in all diagrams shown in Fig. \ref{RCPQa}, the scalar curvature is positive in the range of $0<v<v_{0}=2$ that implies a repulsive interaction between the microscopic black hole molecules. The observed change in the sign of $R$ might be explained by the quantum effects (can be created because of the logarithmic term appeared in entropy) which are dominated in the region $0<v<2$. Therefore, in this region  microscopic molecules strongly tend to interact repulsively with each other, whereas in the region $v>2$ by decreasing quantum effects, interaction between molecules will be attractive and repulsive interactions become weaker. Furthermore, near $v=2$ we observe a balance between repulsive and attractive interactions which is characterized by $R=0$. Contrary to our expectation, by vanishing the Gauss Bonnet coupling $\alpha$ the quantum effect still remain in the scalar curvature. It means that a quantum statistical property is not left in the scalar curvature, even in  $\alpha \to 0$ limit. For ideal quantum gases obeying Gentile's statistics, one can observe a similar phenomena. In spite of the fact that $R$ is zero for ideal classical gases, the scalar curvature has non-zero values for Bose–Einstein (BE) and Fermi–Dirac (FD) statistics in the classical limit \cite{reffff9}. 

 It is worthwhile noting that since we expand the thermodynamic manifold into the four dimensional space such that one of the dimensions is related to $\alpha$ coefficient, the quantum effects of 4D GB-AdS black holes will be important. On the other hand, in previous section one could not observe these quantum effects because thermodynamic manifold was embedded in two dimensional space and $\alpha$ coefficient was presumed to be a constant not as a dimension of the manifold.   

\begin{figure}[tbp]
\includegraphics[scale=0.5]{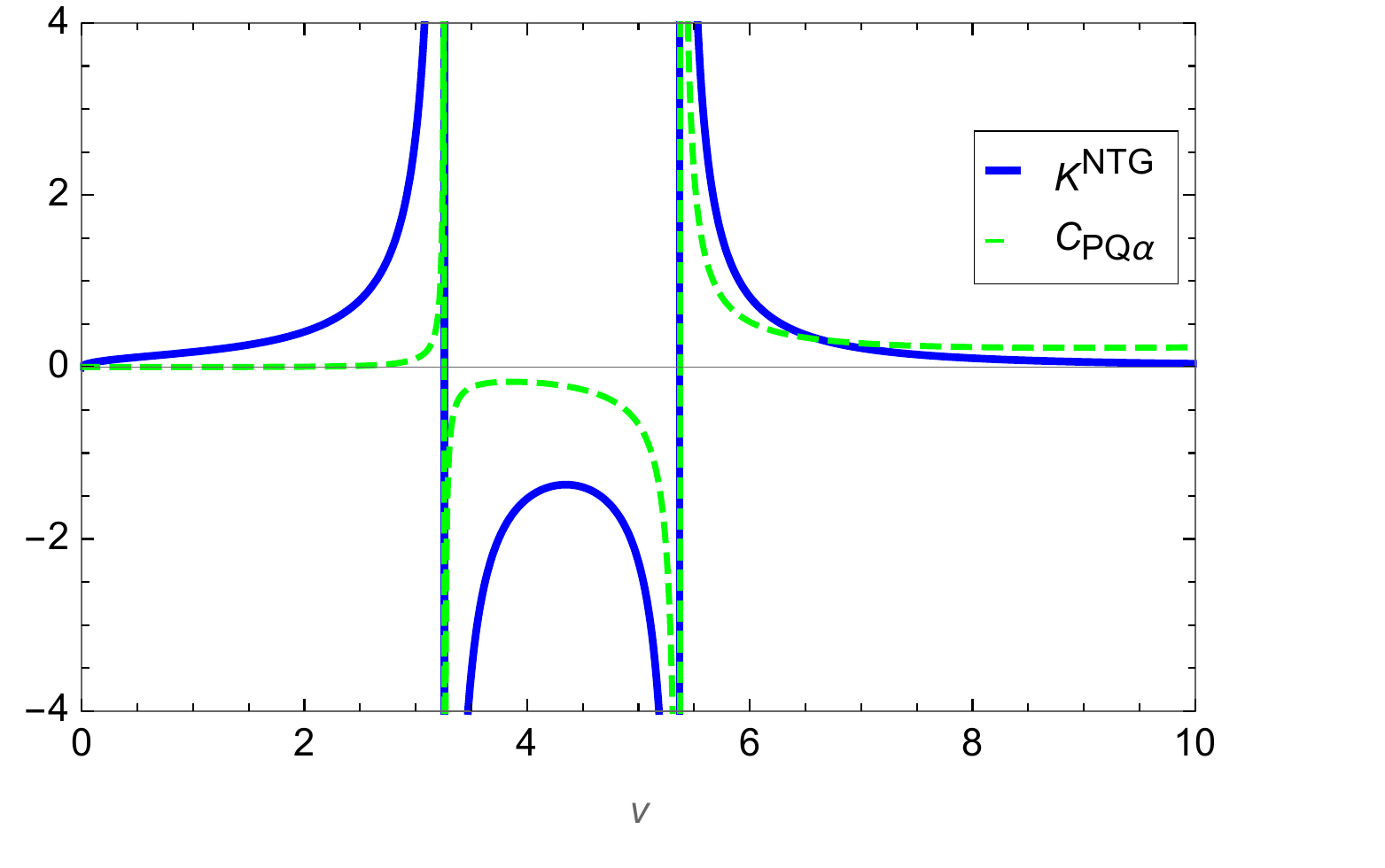}
\includegraphics[scale=0.5]{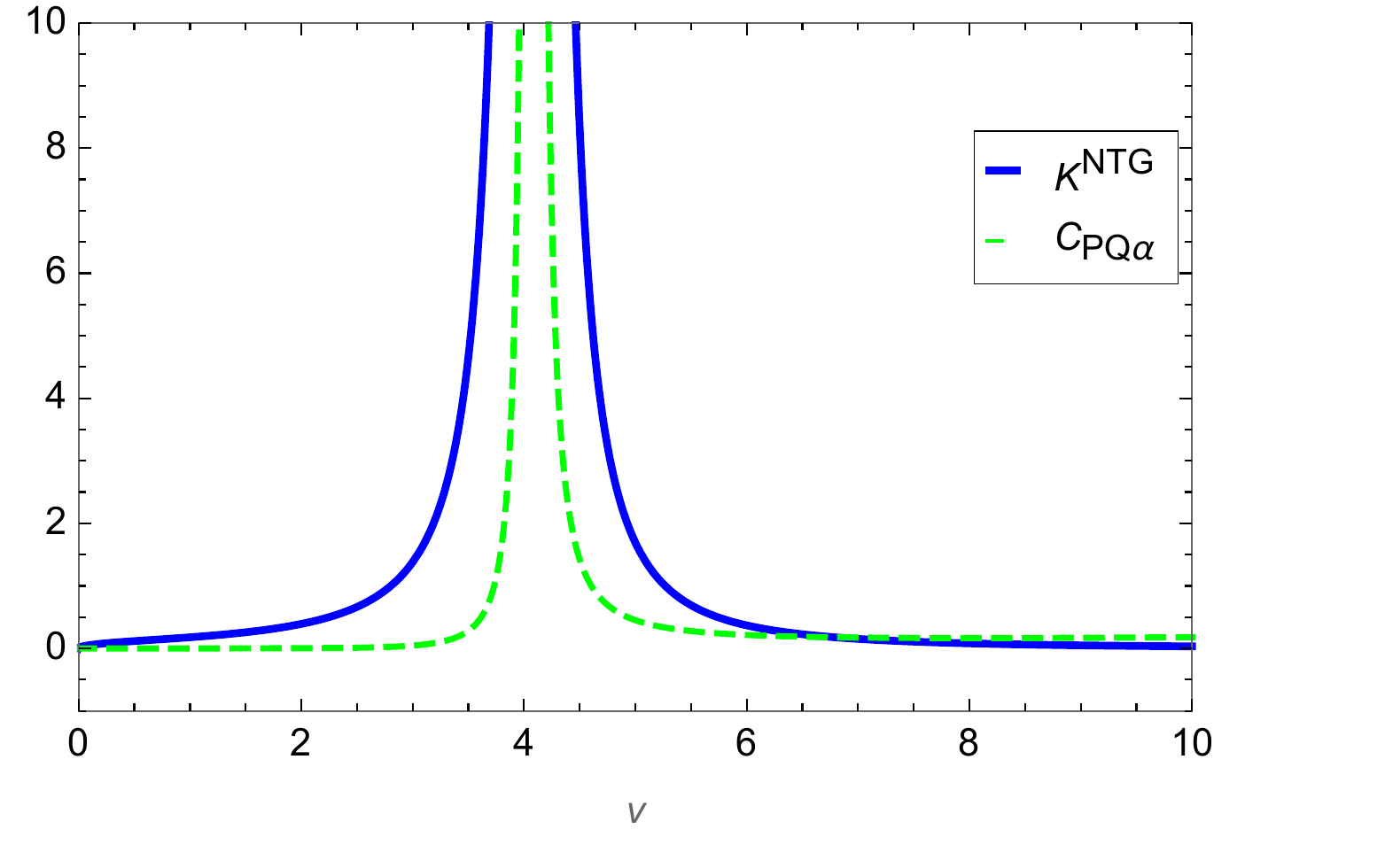}
\begin{center}
\includegraphics[scale=0.5]{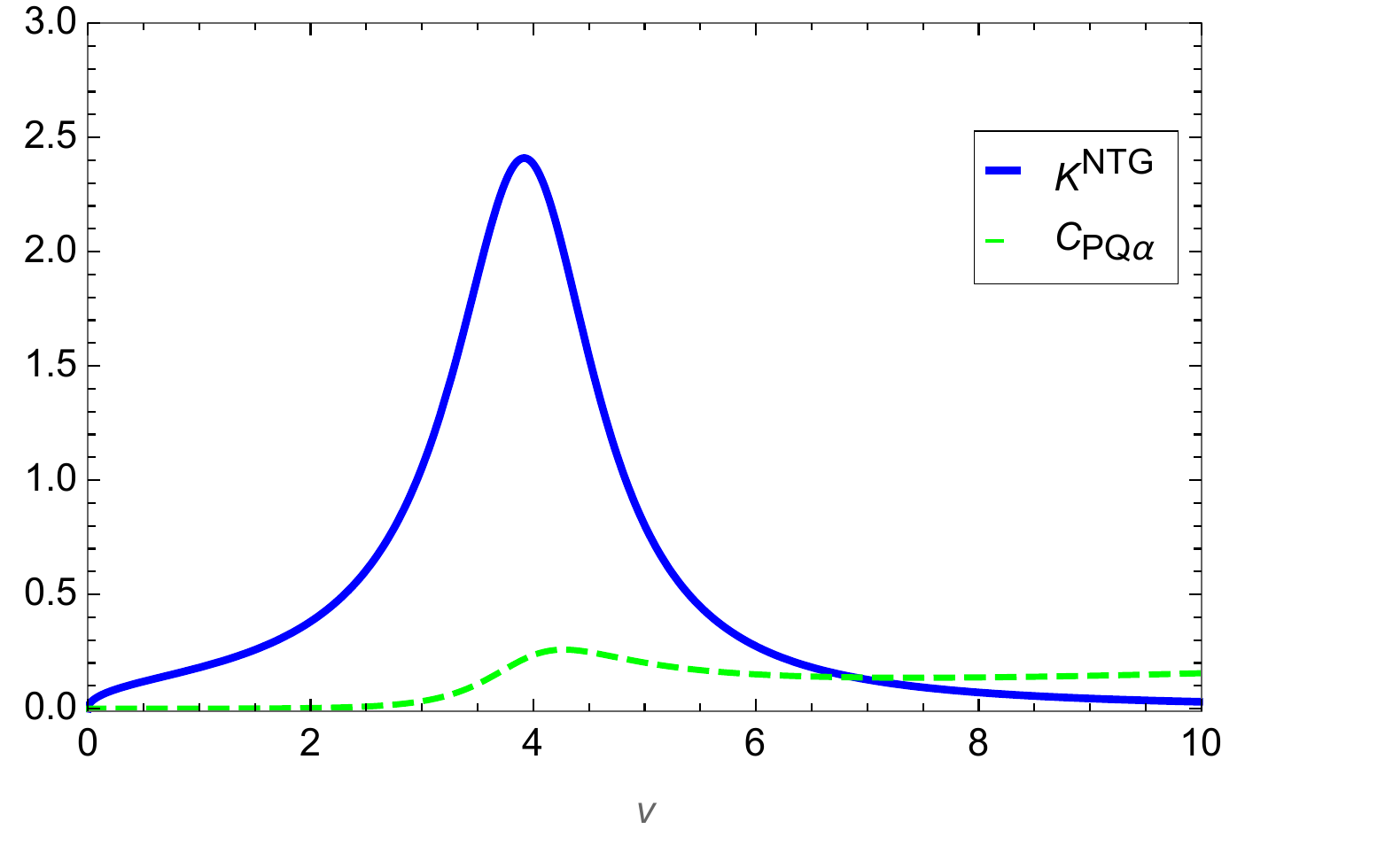}
\end{center}
\caption{The diagram of the specific heat ($\times 10^{6}$) and extrinsic curvature versus the specific volume $v$ for a 4D GB- AdS black hole. Form left to right we consider $T=\{0.0444,0.0474,0.0494\}$ where $T_c=0.0474$ for constant values of the electric charge $Q=0.5$
and Gauss-Bonnet coupling $\alpha=0.2$. Here, we have taken $v_{0}=2$. \label{KCPQa}}
\end{figure}
Allow us now to analyze the nature of the phase transition using thermodynamic hypersurfaces in lower dimensions. To do this, we force ourselves to sit down on the constant
$Q$ hypersurface with the orthogonal normal vector,
\begin{equation}
n_{Q}=\frac{1}{\sqrt{|g^{QQ}|}}=\sqrt{\frac{2}{Tv}}.
\end{equation}
Therefore, the extrinsic curvature is given by
\begin{equation}
K^{NTG}=\frac{4 Q \sqrt{2 T v}}{8 Q^2+v^2(\pi T v-1)+8 \alpha (1+3 \pi T v)}
\end{equation}
Interestingly, it diverges at 
 phase transition points and exhibit a similar sign behavior like the heat capacity around such points as illustrated in Fig. \ref{KCPQa}.  
 
  In a neutral GB-AdS black hole case, we also need to consider the $Q$- zero hapersurface in thermodynamic space. It should be noted that, geometrically setting $Q$ to zero is equivalent to sitting on the constant $Q$ hypersurface ($Q$- zero hapersurface). Taking advantage of Eq. (\ref{metB}), the metric elements induced on this hypersurface are obtained to be
\begin{equation}
 \hat{g}^{in}=\left( \begin{matrix}
 0 & 0  & -\frac{4 \pi}{T} \ln \Big(\frac{v}{v_{0}}\Big)\\
 0& -\frac{v^2 (\pi T v-1)+8\alpha (1+3 \pi T v)}{2 T v^3} &0 \\
- \frac{4 \pi }{T} \ln \Big(\frac{v}{v_{0}}\Big)&0&0\\
 \end{matrix}\right)
 \end{equation}
Therefore, the intrinsic curvature $R^{in}$ of $Q$-zero hypersurface reads
\begin{equation}
R^{in}=\frac{\pi T v(v^2-\pi T v^3-8 \alpha-24 \pi T v \alpha)-2 (v^2(-1+2 \pi T v (2+\pi T v)+8\alpha (1+2 \pi T v)))\ln\Big(\frac{v}{v_{0}}\Big)}{\pi \Big(v^2 (\pi T v-1)+8 \alpha(1+3 \pi T v)\Big)^2\ln\Big(\frac{v}{v_{0}}\Big)^2 }
\end{equation}
It is worthwhile noted that the Ricci scalar in a four dimensional space is related to the above curvature in the three dimensional space  via Gauss-Codazzi relation. Before discussing about the case study, it seems useful to identify the critical point for one neutral GB-AdS black hole.  With the help of the equation of state Eq. (\ref{P1}), the critical point in a neutral GB-AdS is given by
 \begin{equation}
 T_{c}=\frac{\sqrt{2\sqrt{3}-3}}{6 \pi \sqrt{2 \alpha}} \hspace{1cm} v_{c}=2 \sqrt{2 \alpha} \sqrt{3+2 \sqrt{3}} \hspace{1cm} P_{c}=\frac{15-8 \sqrt{3}}{288 \pi \alpha}
 \end{equation}
 and the ratio,
 \begin{equation}
 \frac{P_{c} v_{c}}{T_{c}}=\frac{1}{12}(6-\sqrt{3})
 \end{equation}
 is slightly smaller than the van-der Waals ratio $3/8$ \cite{Kubiznak:2012wp}.
In the reduced parameter space, the equation of state (\ref{P1}) has the following form
\begin{equation}
\hat{P}=(9-4 \sqrt{3})\frac{1}{11 \hat{v}^4}+4(3 \sqrt{3}-4) \frac{\hat{T}}{11 \hat{v}^3}-6(1+2 \sqrt{3})\frac{1}{11 \hat{v}^2}+4(6+\sqrt{3}) \frac{\hat{T}}{11 \hat{v}}
\end{equation}
where the reduced pressure, temperature, and specific volume are defined by $\hat{P}=\frac{P}{P_{c}}$, $\hat{T}=\frac{T}{T_{c}}$, and $\hat{v}=\frac{v}{v_{c}}$, respectively. More interestingly, this reduced state equation does not depend on $\alpha$ parameter. Now one can identify the spinodal curve which satisfies $(\partial_{\hat{v}} \hat{P})_{\hat{T}}=0$. In fact, this curve separates the metastable phase from the unstable phase. In addition, heat capacity and the scalar curvature diverge along this curve. Solving mentioned conditions, spinodal curve has a below compact form
\begin{equation}
T_{sp}=\frac{3(2+\sqrt{3}) \hat{v}^2-\sqrt{3}}{3 \hat{v}+(3+2 \sqrt{3}) \hat{v}^3}
\end{equation}
where $\frac{\sqrt{2-\sqrt{3}}}{\sqrt[4]{3}}<\hat{v}<1$ is for the small black hole spinodal curve, while $\hat{v}>1$ is for the large black hole spinodal curve (see left had side of Fig. \ref{Spin}) \footnote{
Because of the logarithmic correction term appeared in entropy, it may be hard to obtain an analytical expression for the coexistence curve for 4D Gauss-Bonnet AdS black holes.}.  
\begin{figure}[tph]\label{Spin}
\includegraphics[scale=0.5]{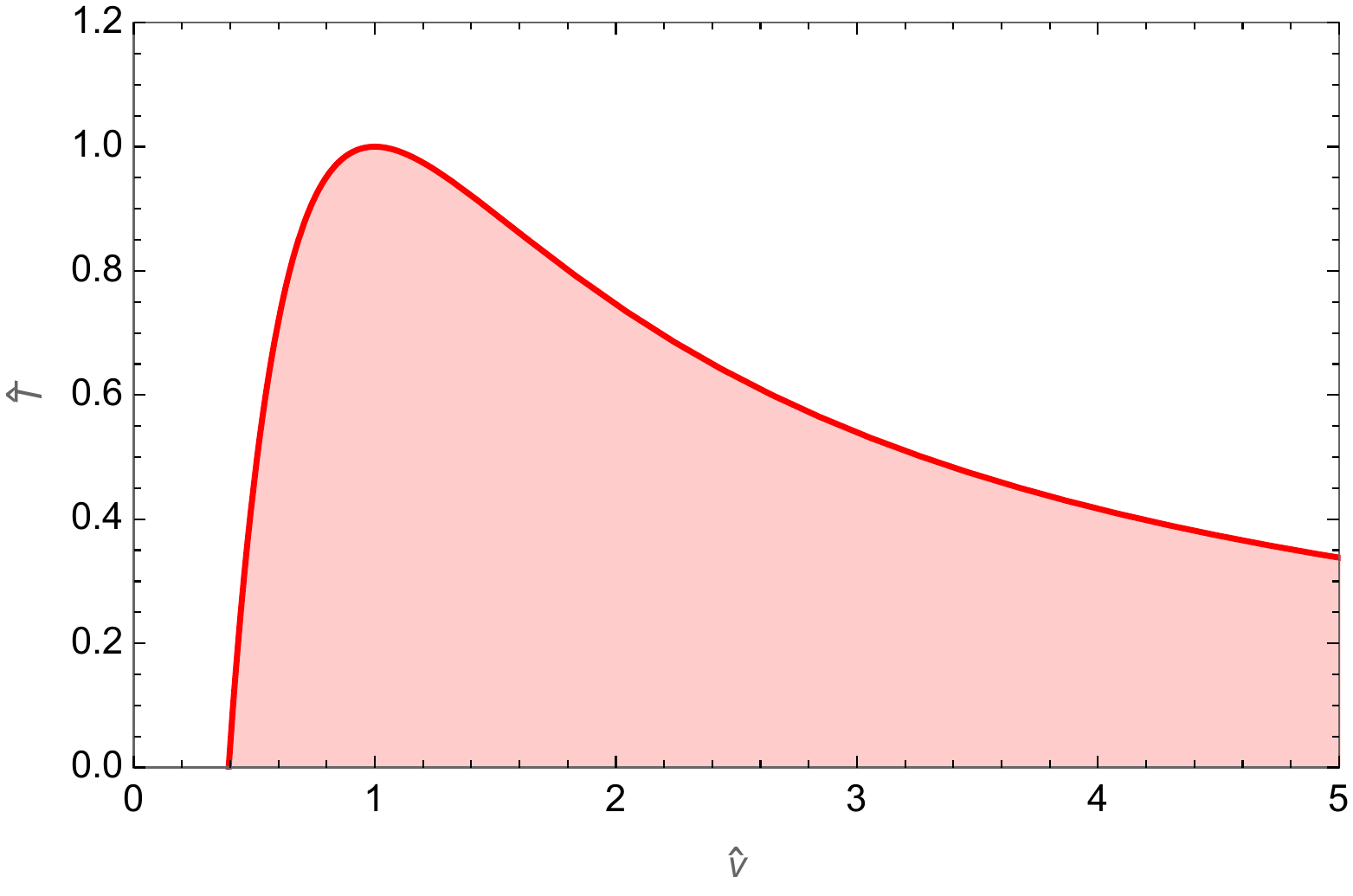}
\includegraphics[scale=0.5]{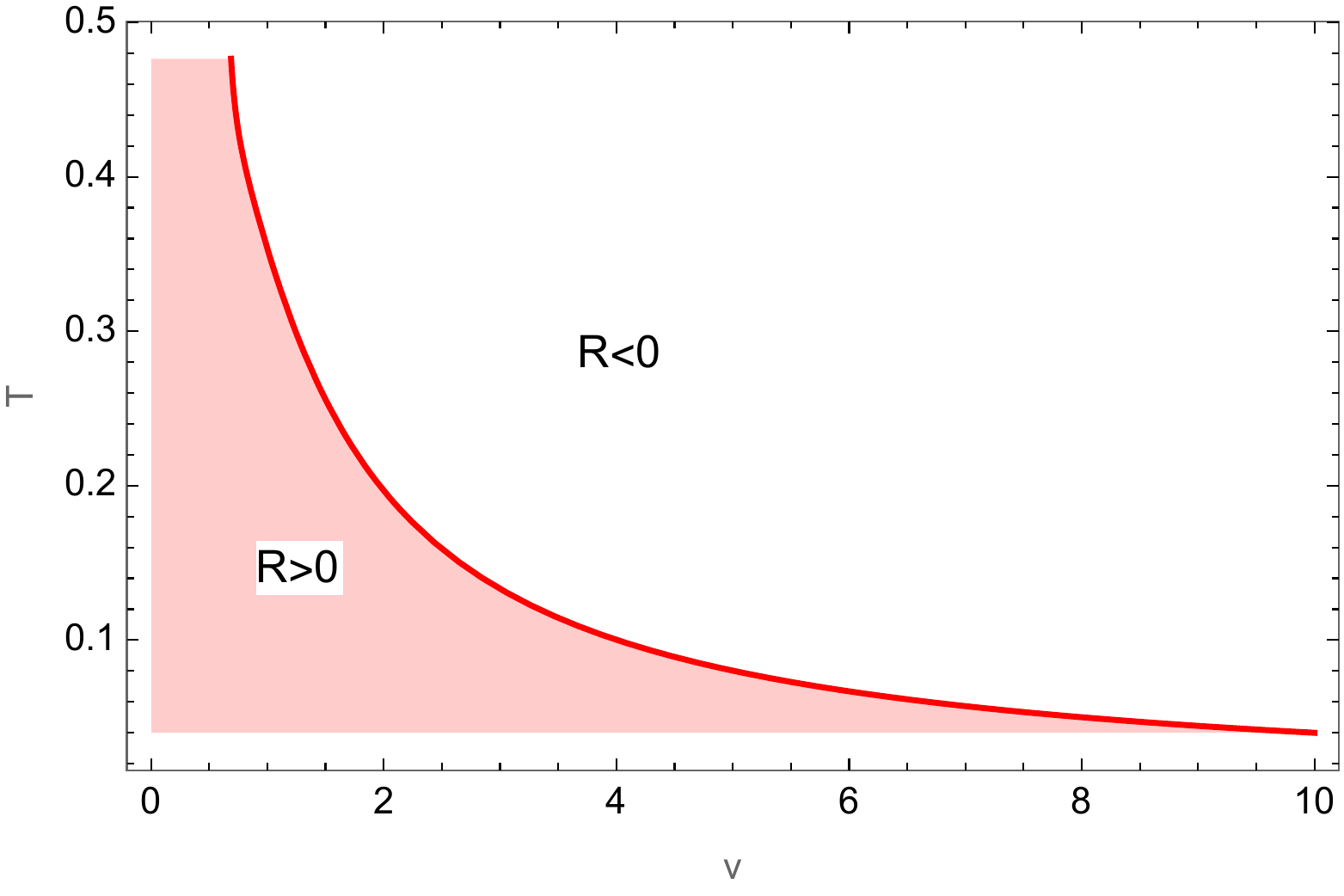}
\caption{Left: The spinodal curve of the four-dimensional neutral GB-AdS black hole, Right: The sign-changing curve where the scalar
curvature $R^{in}$ changes its sign. Here, we have taken $v_{0}=2$ and $\alpha=0.2$.\label{Spin}}
\end{figure}
\begin{figure}[tbp]
\includegraphics[scale=0.5]{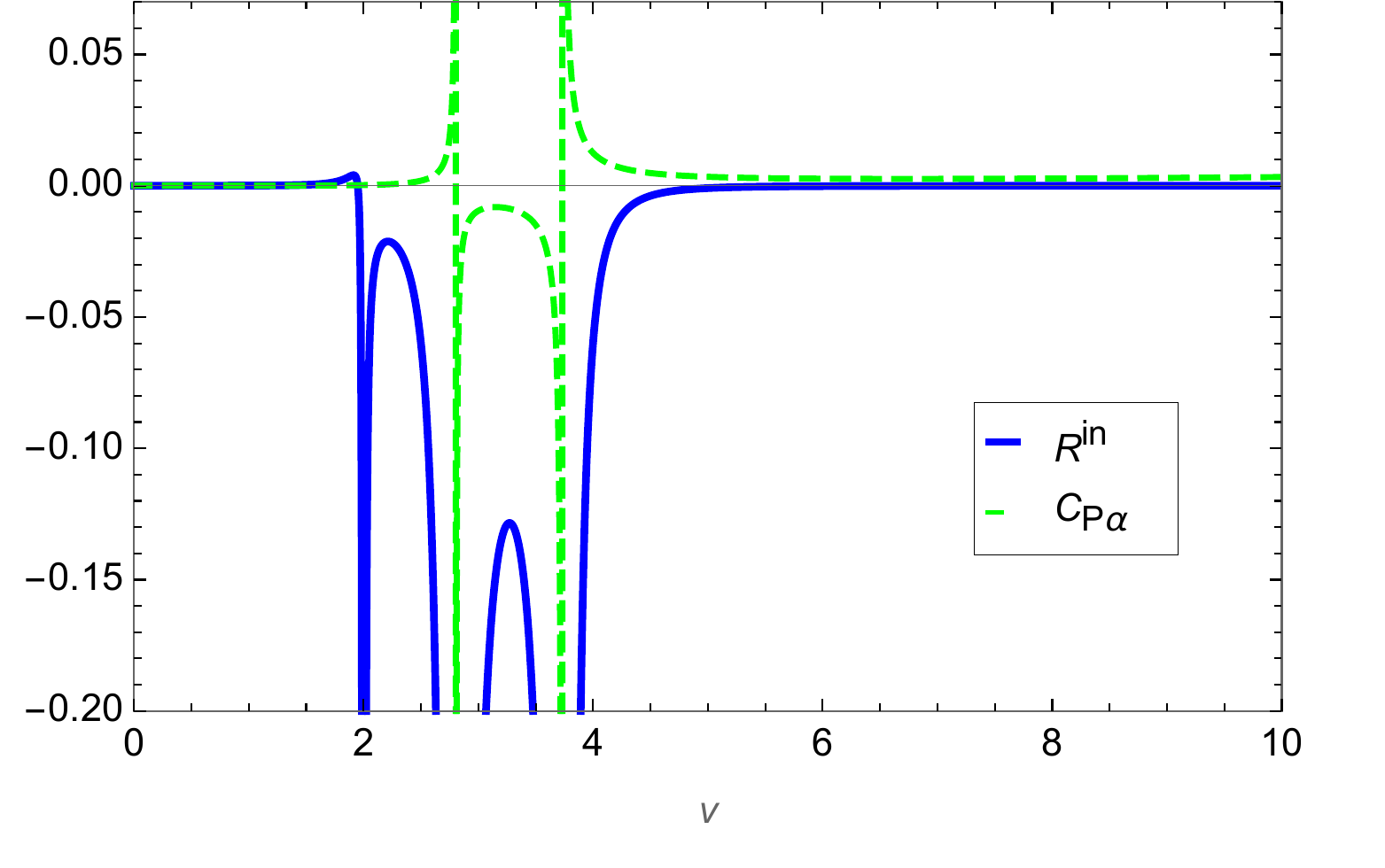}
\includegraphics[scale=0.5]{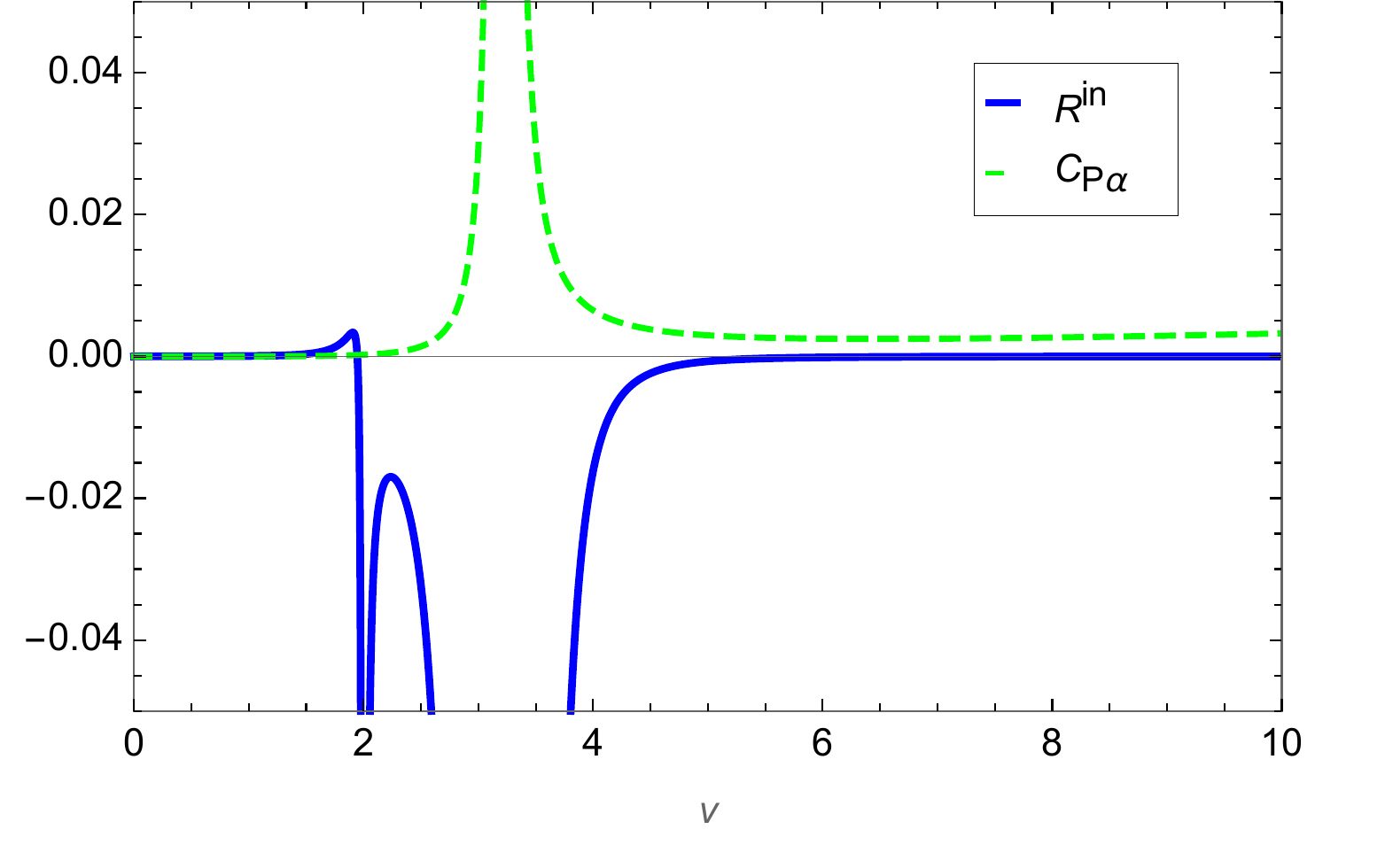}
\begin{center}
\includegraphics[scale=0.5]{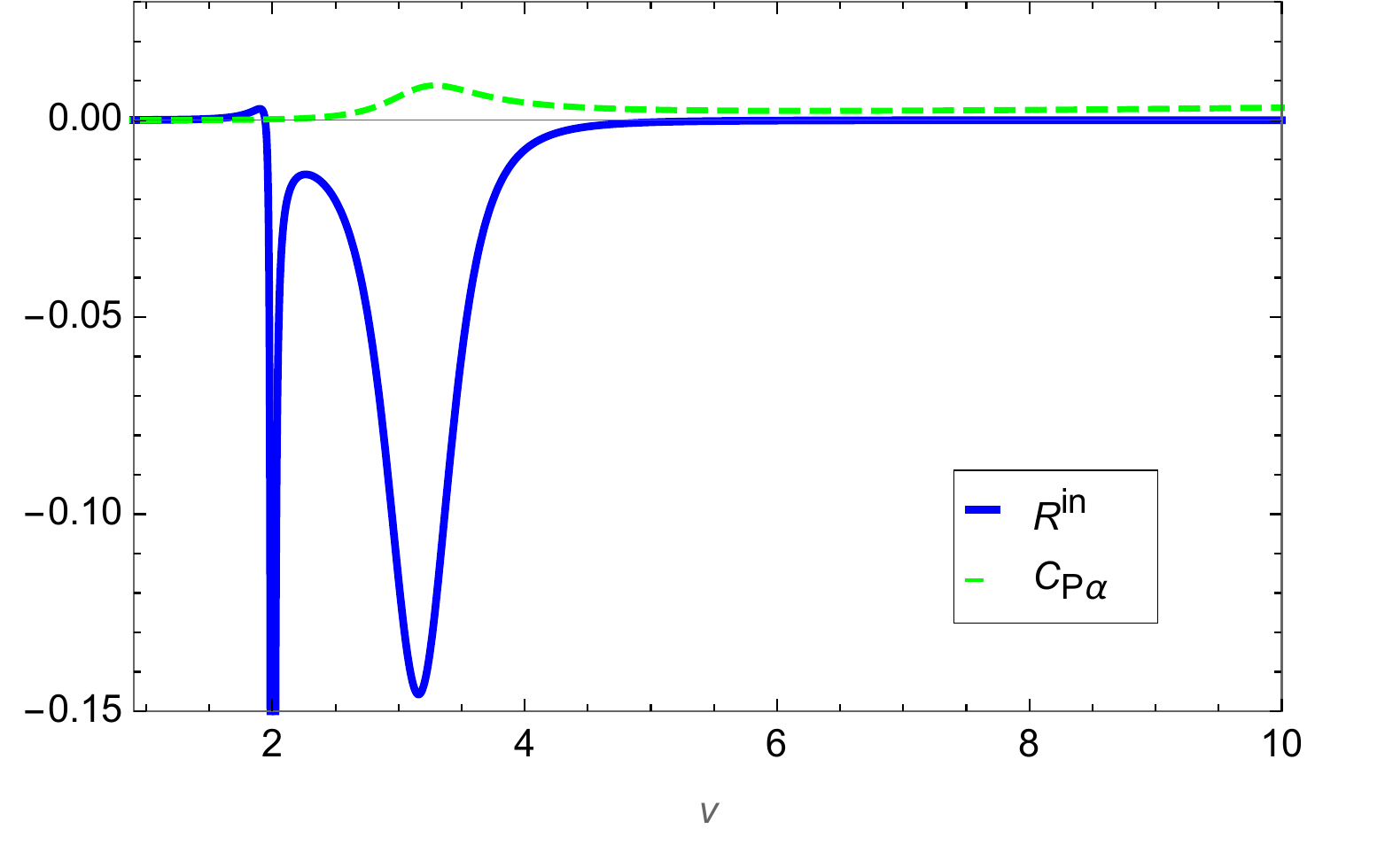}
\end{center}
\caption{The diagram of the specific heat ($\times 10^{5}$) and intrinsic curvature ($\times 10^{4}$) versus the specific volume $v$ for a 4D GB- AdS black hole. Form left to right we consider $T=\{0.0561,0.0571,0.0581\}$ where $T_c=0.0571$ for a constant value of 
the Gauss-Bonnet coupling $\alpha=0.2$. Here, we have taken $v_{0}=2$.\label{Rin}}
\end{figure}

  In Fig. \ref{Rin}, we have depicted the behavior of $R^{in}$ and $C_{P, \alpha}=C_{P,Q,\alpha}(Q=0)$ for $T>T_{c}$, $T=T_{c}$ and $T<T_{c}$, respectively. When $T>T_{c}$, we observe three negative divergent
points where the first happens at $v=v_{0}=2$ generated from $\ln (v/v_{0})$ in denominator of curvature. Moreover, the other two points get closer by increasing $T$. At the critical temperature $T=T_{c}=0.0571$, these two divergent points merge to form a single divergence at $v=v_{c}=3.215$. More importantly, in the presence of quantum effects, we observe a positive $R^{in}$ near the region $0<v<v_{0}$, which indicates a repulsive interaction among microstructures of the black hole.
The right hand side of Fig. \ref{Spin} presents the sign-changing curve of $R^{in}$. Below this curve, $R^{in}$ is positive, whereas above it, $R^{in}$ will be negative. Clearly,  by increasing the temperature value, this positive region shown in Fig. \ref{Rin}, disappears and the scalar curvature takes the negative values. A reasonable explanation for this result may be that the quantum effects disappear when we increase temperature. Nevertheless, the positive $R^{in}$ can not excluded at low temperature for small black holes. This demonstrates that the repulsive interaction dominates among the microstructures for the small black hole with low temperature in the 4D neutral GB-AdS black hole case. This is in contrast to the five dimensional (uncharged) case where the attractive interaction is dominant in black hole microstructures (see appendix \ref{appc}) \footnote{The same result is also confirmed by the authors in Ref. \cite{Wei:2020poh} for the 4D  neutral GB-AdS black hole in comparison with a 5D GB-AdS black hole \cite{GW1} along the consistence curve for large and small black holes. Geometrically, their findings can be reproduced by considering a $\alpha$-constant hypersurface in $Q$-zero manifold (hypersurface). This results in the fact that the quantum effects disappear.}. The most remarkable result to emerge from Appendix \ref{appc} for 5D neutral GB-AdS black holes is that an attractive interaction exists only among the microstructures of low temperature small black holes. Despite this, repulsive interactions can be significant in the high temperature small black hole case.

\section{Conclusions}\label{sec5}
The purpose of the current study was to apply the NTG geometry to determining phase structure of the 4D GB-AdS black hole in both neutral and charged cases.   In this paper, we have presented a general form of NTG geometry for other thermodynamic potentials which are made from Legendre transformations. This novel geometry provides us with a one-to-one correspondence between heat capacity phase transitions and curvature singularities. In order to understand the critical behavior of the 4D GB-AdS black hole, in the first step we have used NTG geometry for the normal phase space when neither the cosmological constant $\Lambda$ (or $l$) nor the Gauss Bonnet coefficient $\alpha$ have  variations in the first law of thermodynamics. By taking advantage of the extrinsic curvature for a special type of the hypersurface immersed in the normal phase space, we have found that the extrinsic curvature not only shows the precise location of the phase transition point but also has the same sign as heat capacity around that point. 

For the next step, we focused on the extend phase space when the variation of cosmological constant can be interpreted as thermodynamic pressure in the first law of thermodynamics. In this regard, we have calculated the scalar curvature for 4D  charged GB-AdS black holes. What we were surprised to find is the fact that an extra singular point at $v=2$ resulted in failing the one-to-one correspondence between scalar curvature singularities and phase transition points. The reason for this contradictory result is a consequence of the existence of the logarithmic term, which is describing the quantum effects, in 4D GB-AdS black hole cases. 

From geometrical point of view, the thermodynamic phase space for a neutral 4D GB-AdS black hole should be limited to the $Q$-zero hypersurface immersed in the thermodynamic manifold for a 4D  charged GB-AdS black hole. The intrinsic curvature of this hypersurface provides considerable insight into the phase structure and microstructure of a 4D neutral GB-AdS black hole. 

As mentioned in the Introduction, the kind of interaction between black hole molecules can be found from the sign of the thermodynamic curvature. Therefore, we observed that there was a positive intrinsic curvature $R^{in}$ in volume range of $0<v<2$, which indicates a repulsive interaction among microstructures of the black hole.
At high temperature limit, this area disappears and scalar curvature always takes negative values, which implies that only attractive interaction exists among the microstructures. Nevertheless, at low temperature for small black holes $R^{in}$ still has positive value. It means that repulsive interactions dominate between micromolecules of low temperature small black holes. These finding are in contradiction with results found for the five dimensional (uncharged) case where the attractive interaction is dominant in small black hole microstructures at low temperatures. 


\section*{Acknowledgments}

We would like to thank Behroz Mirza and Mohammad Ali Gorji for valuable suggestions and discussions. We gratefully acknowledge constructive comments of anonymous referees. 
\appendix
\section{ Bracket notation and Partial derivative}\label{A1}
Partial derivative of the functions depending on $n+1$ independent variables can be defined by \cite{reff13},
\begin{equation}
{{\left( \frac{\partial f}{\partial g} \right)}_{{{h}_{1}},.....,{{h}_{n}}}}=\frac{{{\left\{ f,{{h}_{1}},...,{{h}_{n}} \right\}}_{{{q}_{1}},{{q}_{2}},...,{{q}_{n+1}}}}}{{{\left\{ g,{{h}_{1}},...,{{h}_{n}} \right\}}_{{{q}_{1}},{{q}_{2}},...,{{q}_{n+1}}}}}
\end{equation}
where all of $f$, $g$, and ${{h}_{n}}$ ($n=1,\,2,\,3,\,...$) are functions of ${{q}_{i}},i=1,...,n+1$ variables and $\{.,.,.\}$ denotes Nambu bracket which is defined as,
\begin{eqnarray}
{{\left\{ f,{{h}_{1}},...,{{h}_{n}} \right\}}_{{{q}_{1}},{{q}_{2}},...,{{q}_{n+1}}}}=\\
 \nonumber \sum\limits_{ijk....l=1}^{n+1}{{{\varepsilon }_{ijk...l}}\frac{\partial f}{\partial {{q}_{i}}}}\frac{\partial {{h}_{1}}}{\partial {{q}_{j}}}\frac{\partial {{h}_{2}}}{\partial {{q}_{k}}}...\frac{\partial {{h}_{n}}}{\partial {{q}_{l}}}
\end{eqnarray}
In a simple case, when we consider separately $f$,  $g$, and $h$ as a function of ($a, b$), the above formula reduces to 
\be
{{\left( \frac{\partial f}{\partial g} \right)}_{h}}=\frac{{{\left\{ f,h \right\}}_{a,b}}}{{{\left\{ g,h \right\}}_{a,b}}}
\label{N15}
\ee
where $\{.,.\}$ is Poisson bracket which is given by
\be
{{\{f,h\}}_{a,b}}={{\left( \frac{\partial f}{\partial a} \right)}_{b}}{{\left( \frac{\partial h}{\partial b} \right)}_{a}}-{{\left( \frac{\partial f}{\partial b} \right)}_{a}}{{\left( \frac{\partial h}{\partial a} \right)}_{b}}
\ee
\section{Some of the other heat capacities}\label{appB}
In this appendix, we examine phase transition signals of some  specific heats such as,
\begin{eqnarray}
C_{V,Q,\alpha}&=&T\Big(\frac{\partial S}{\partial T}\Big)_{V,Q,\alpha}=T \frac{\{S,V,Q,\alpha\}_{T,v,Q,\alpha}}{\{T,V,Q,\alpha\}_{T,v,Q,\alpha}}=0
\end{eqnarray}
and
\begin{eqnarray}
C_{P,\Phi,\mathcal{A}}&=&T\Big(\frac{\partial S}{\partial T}\Big)_{P,\Phi,\mathcal{A}}
=T \frac{\{S,P,\Phi,\mathcal{A}\}_{T,v,Q,\alpha}}{\{T,P,\Phi,\mathcal{A}\}_{T,v,Q,\alpha}}\\
\nonumber &=& -\frac{4 \pi^2 T \ln\Big(\frac{v}{v_{0}}\Big)\Big((1+4 \pi T v)(v^2+8\alpha)+2 (4 Q^2+v^2 (\pi T v-1)+8\alpha(1+3 \pi T v) )\ln \Big(\frac{v}{v_{0}}\Big)\Big)}{ (1+4 \pi T v)^2}
\end{eqnarray}
We find that there is no phase transition for the above
heat capacities. This result can also be confirmed by using NTG geometry. To do this, let us substitute $\Xi=F=E-TS=M-PV-TS$ with coordinate $X^{i}=(T,V,Q,\alpha)$ into Eq. (\ref{Ru1}), thus we have
\begin{equation}
g^{NTG}_{F}=\frac{1}{T}\left(\begin{matrix}
-F_{TT}& 0& 0 & 0\\
0& F_{VV}& F_{VQ} & F_{V \alpha}\\
0& F_{QV}& F_{QQ} & F_{Q \alpha}\\
0& F_{\alpha V}& F_{\alpha Q} & F_{\alpha \alpha}\\
\end{matrix}\right)=\left(\begin{matrix}
\frac{C_{V,Q,\alpha}}{T^2}& 0& 0 & 0\\
0& \frac{2(3 Q^2+v^2(\pi T v-1)+8\alpha (1+3 \pi T v))}{\pi^2 T v^7}& -\frac{4 Q}{\pi T v^4} & -\frac{2+8 \pi T v}{\pi T v^4}\\
0&-\frac{4 Q}{\pi T v^4} & \frac{2}{Tv} & 0\\
0& -\frac{2+8 \pi T v}{\pi T v^4}& 0 & 0\\
\end{matrix}\right)
\end{equation}
Since the vanishing heat capacity $C_{V,Q,\alpha}$  leads to $g_{TT} \to 0$ or $g^{TT}\to \infty$, the metric is not an invertible matrix. Following Ref. \cite{Wei:2019uqg}, we take $C_{V,Q,\alpha}$ to be a non-zero constant and its value eventually goes to zero \cite{Wei:2019uqg}.
Therefore, we can obtain a normalized scalar curvature $R_{N}$ as
\begin{equation}
R_{N}^{NTG}=C_{V,Q,\alpha} R^{NTG}=-\frac{3+8 \pi T v(2+\pi T v)}{(1+4 \pi T v)^2}
\end{equation}
Analogous with a charged AdS black hole, this normalized scalar curvature is independent of $\alpha$. Clearly, we observe $R$ is always negative, which implies an attractive interaction among the microstructures of a black hole system. Moreover, for a constant $T$ hypersurface with normal vector $n_{T}=\sqrt{\frac{C_{V,Q,\alpha}}{T}}$, the extrinsic curvature is obtained to be
\begin{equation}
K_{N}^{NTG}=\sqrt{C_{V,Q,\alpha}} K^{NTG}=-\frac{3+4 \pi T v}{(1+4 \pi T v)}
\end{equation}
which shows that the phase transition does not occur at any point. 

\section{5D neutral Gauss-Bonnet AdS black hole}\label{appc}
\begin{figure}[tbp]
 	\includegraphics[scale=0.5]{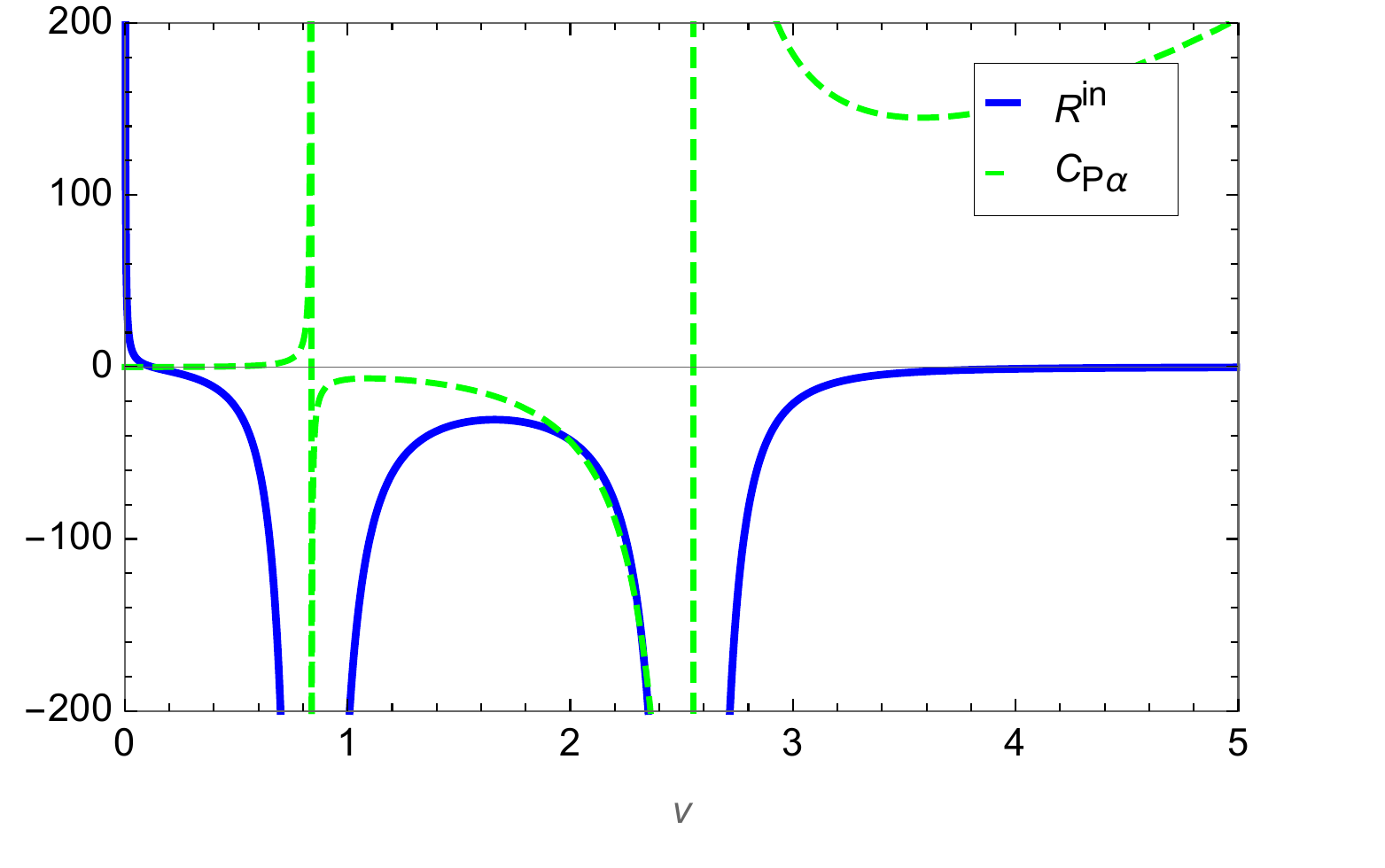}
 	\includegraphics[scale=0.5]{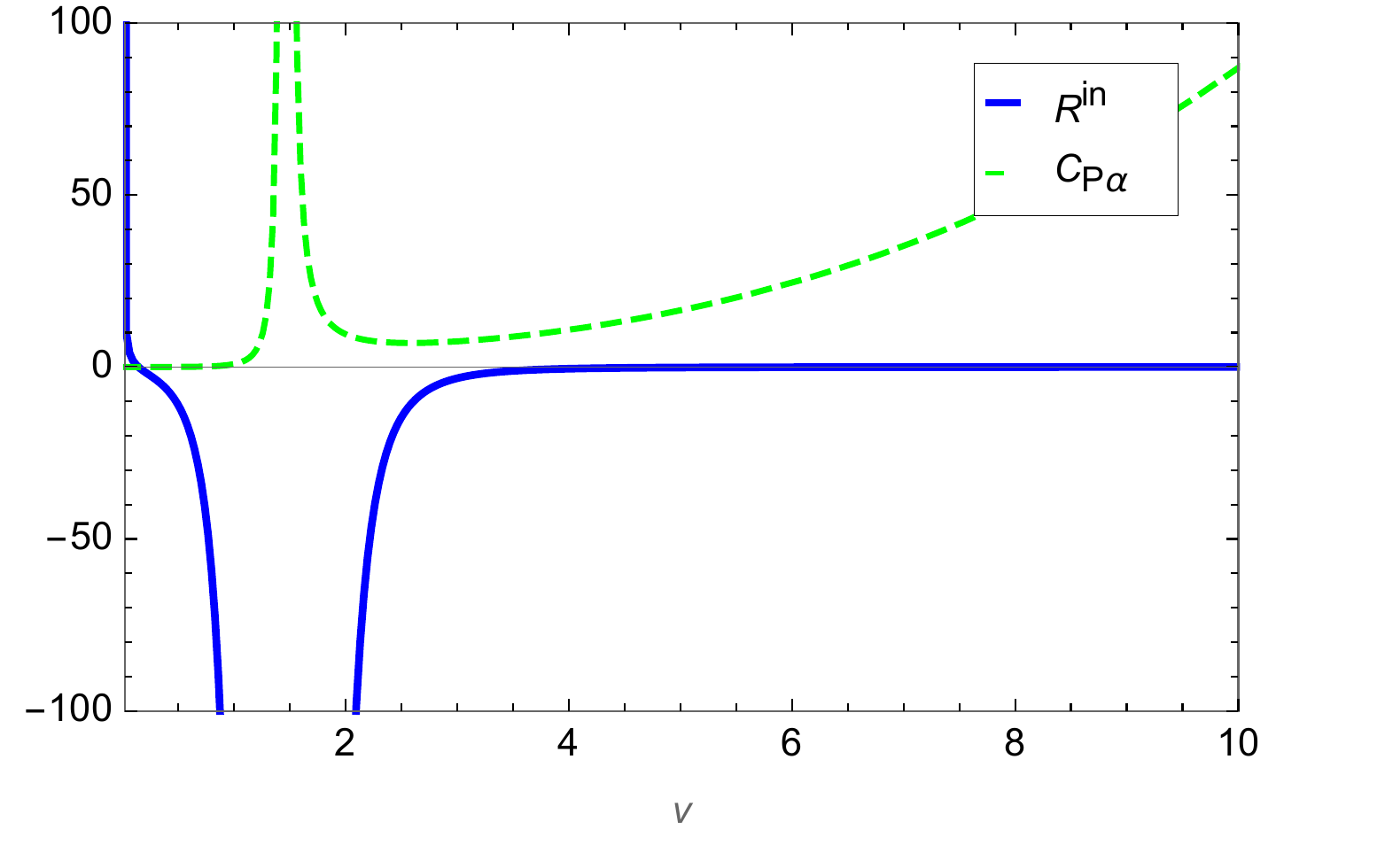}
 	\begin{center}
 		\includegraphics[scale=0.5]{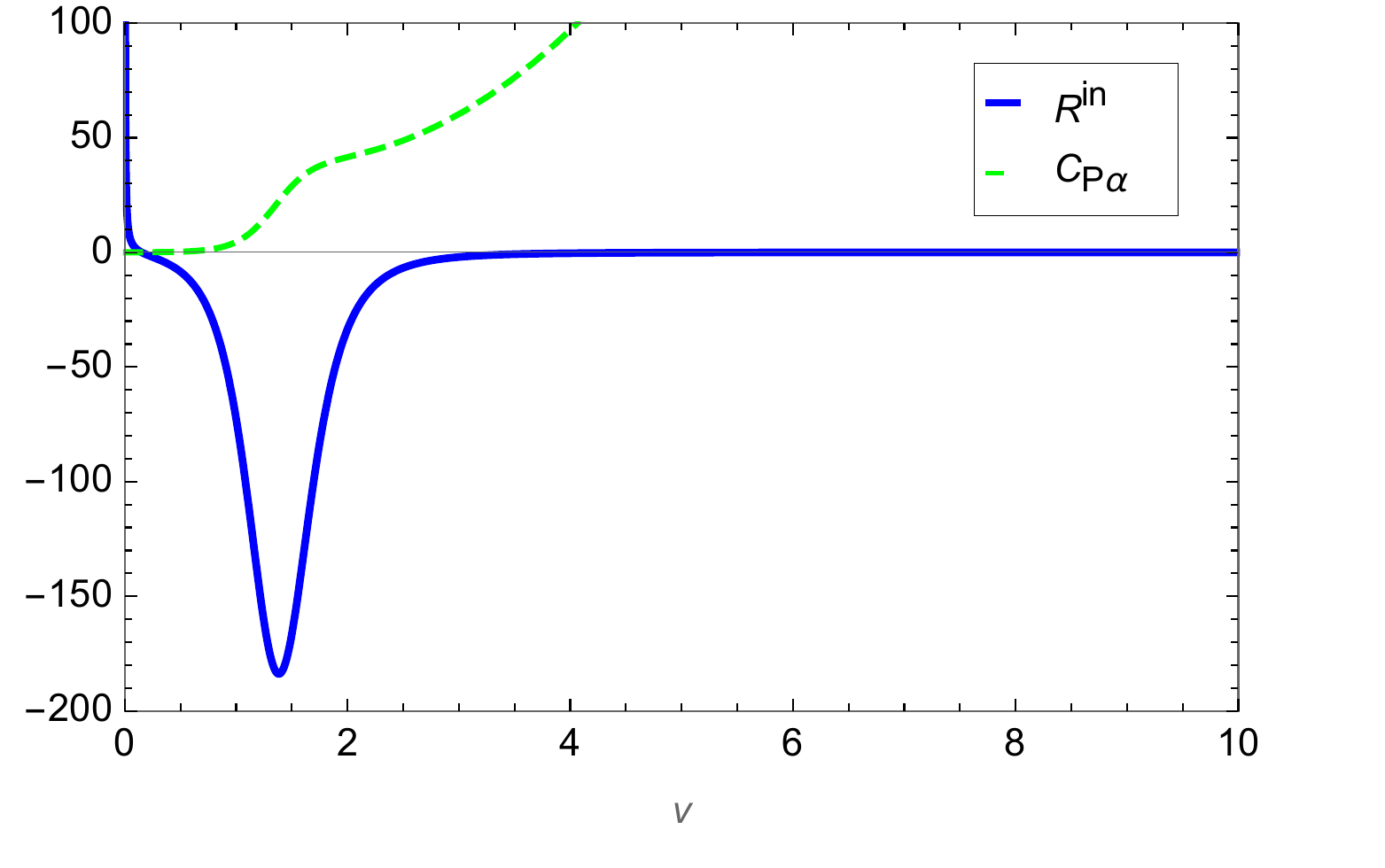}
 	\end{center}
 	\caption{The diagram of the specific heat ($\times 10^{2}$) and scalar curvature ($\times 10^{-1}$) versus the specific volume $v$ for a 5D GB- AdS black hole. Form left to right we consider $T=\{0.1252,0.1452,0.1552\}$ where $T_c=0.1425$ for constant values of the Gauss-Bonnet coupling $\alpha=0.2$. \label{GBRCPQa} }
 \end{figure}

In the five-dimensional neutral GB-AdS black hole case, the metric function takes the below form \cite{Cai:2001dz}. 
\begin{equation}
f=1+\frac{r^2}{2 \alpha} \Big[1-\sqrt{1+ 4 \alpha\left(8 \frac{M}{3 \pi r^4}-\frac{1}{l^2}\right)}\Big]
\end{equation}
By imposing the condition $f(r_{+})=0$ at the horizon, we obtain the mass of this black hole to be
\begin{equation}
	M=\frac{3 \pi}{8} \Big(r_{+}^2+\frac{r_{+}^4}{l^2}+\alpha\Big)
\end{equation}  
The Hawking temperature of the black hole is also given by
\begin{equation}
	T=\frac{f'(r_{+})}{4\pi}=\frac{l^2 r_{+}+2 r_{+}^3}{2 \pi l^2\Big( r_{+}^2+2 \alpha \Big) }
\end{equation}
As mentioned, in the extended phase space by introducing the pressure $P=-\Lambda/8 \pi$, the mass should be treated as the enthalpy potential rather than the internal energy \cite{Kubiznak:2012wp}. Other thermodynamic quantities such as entropy and thermodynamic volume can be obtained through thermodynamic identities. Therefore, it is straightforward to verify that
\begin{eqnarray}
S&=&\frac{3}{8} \pi^2 v (\frac{9}{16} v^2+6 \alpha)\\
V&=& \frac{81}{512} \pi^2 v^4\\
P&=&\frac{T}{v}-\frac{2}{3 \pi v^2}+\frac{32 T \alpha}{9 v^3}
\end{eqnarray} 
where $v=4 r_{+}/3$. The critical point is determined by solving $(\partial_{v}P)_{T,\alpha}=0$ and $(\partial^{2}_{v}P)_{T,\alpha}=0$ at the same time. Thus, one gets \cite{Mo:2016sel}
\begin{equation}
T_{c}=\frac{1}{2 \pi \sqrt{6 \alpha}}; \hspace{0.5cm} v_{c}=\sqrt{\frac{32 \alpha}{3}}
\end{equation}
Note that the heat capacity $C_{P,\alpha}$ which is given by
\begin{equation}
	C_{P, \alpha}=T \Big(\partial_{T}S\Big)_{P,\alpha}= \frac{3 \pi^3 T v (9 v^2+32 \alpha)^2}{128(v(-4 +3 \pi T v)+32 \pi T \alpha)}
\end{equation}
is divergent at the critical point. Let us now use NTG geometry to examine the critical behavior of the heat capacity $C_{P, \alpha}$ and the microstructures of the black hole. Our steps proceed exactly in the same way as what was done in Section \ref{sec4} by ignoring all metric elements associated with the electric charge.  By considering enthalpy $H$ as a thermodynamic potential in NTG metric, we have
\begin{equation}
g^{NTG}_{H}=\frac{1}{T}\left(\begin{matrix}
-H_{SS}& 0 & 0\\
0& H_{PP} & H_{P \alpha}\\
0& H_{\alpha P} & H_{\alpha \alpha}\\
\end{matrix}\right)
\end{equation}
This gives the 
 scalar curvature to be
\begin{equation}
R^{in}=-\frac{64 T \Big(v(20+9 \pi T V)-32 \pi T \alpha\Big)}{27 \pi v \Big(v(-4+3 \pi T v)+32 \pi T \alpha\Big)^2}
\end{equation}
 Fig. \ref{GBRCPQa} shows the behavior of $R^{in}$ and $C_{P, \alpha}$ for $T>T_{c}$, $T=T_{c}$ and $T<T_{c}$, respectively. For $T>T_{c}$, we observe that there are two points at which the scalar curvature diverges. However, these points coincide with each other at $v=v_{c}=1.46059$ for $T=T_{c}$. By increasing temperature, for $T>T_{c}$ this divergent behavior can be vanished. As expected, our finding confirms the one-to-one correspondence between phase transition points and curvature singularities. It is worth mentioning that the quantum effects are absent in 5D GB-AdS case. 
  In the region,
 \begin{equation}
 0<v<\frac{1}{9 \pi T} \Big(2 T \sqrt{\frac{25}{T^2}+72 \pi^2 \alpha}-10\Big)
 \end{equation}
 we also observe a positive values for $R^{in}$. As shown in the right hand side of Fig. \ref{SpinGB}, this region will be vanished at low temperature limit, whereas it shifts to $0<v<v_{c}$ for high temperature limit. More precisely, for small black hole at the low temperature, the positiveness of $R^{in}$ can be excluded and $R^{in}$ is always negative, which shows only attractive interaction exist among the microstructures. As another consequence of the scalar curvature, the spinodal curve can be expressed as
\begin{equation}
	T_{SP}=\frac{2 \hat{v}}{1+\hat{v}^2}
\end{equation}
  which separates the metastable branch from the unstable branch (please see the left hand side of Fig. \ref{SpinGB}).   
\begin{figure}[tph]\label{SpinGB}
	\includegraphics[scale=0.5]{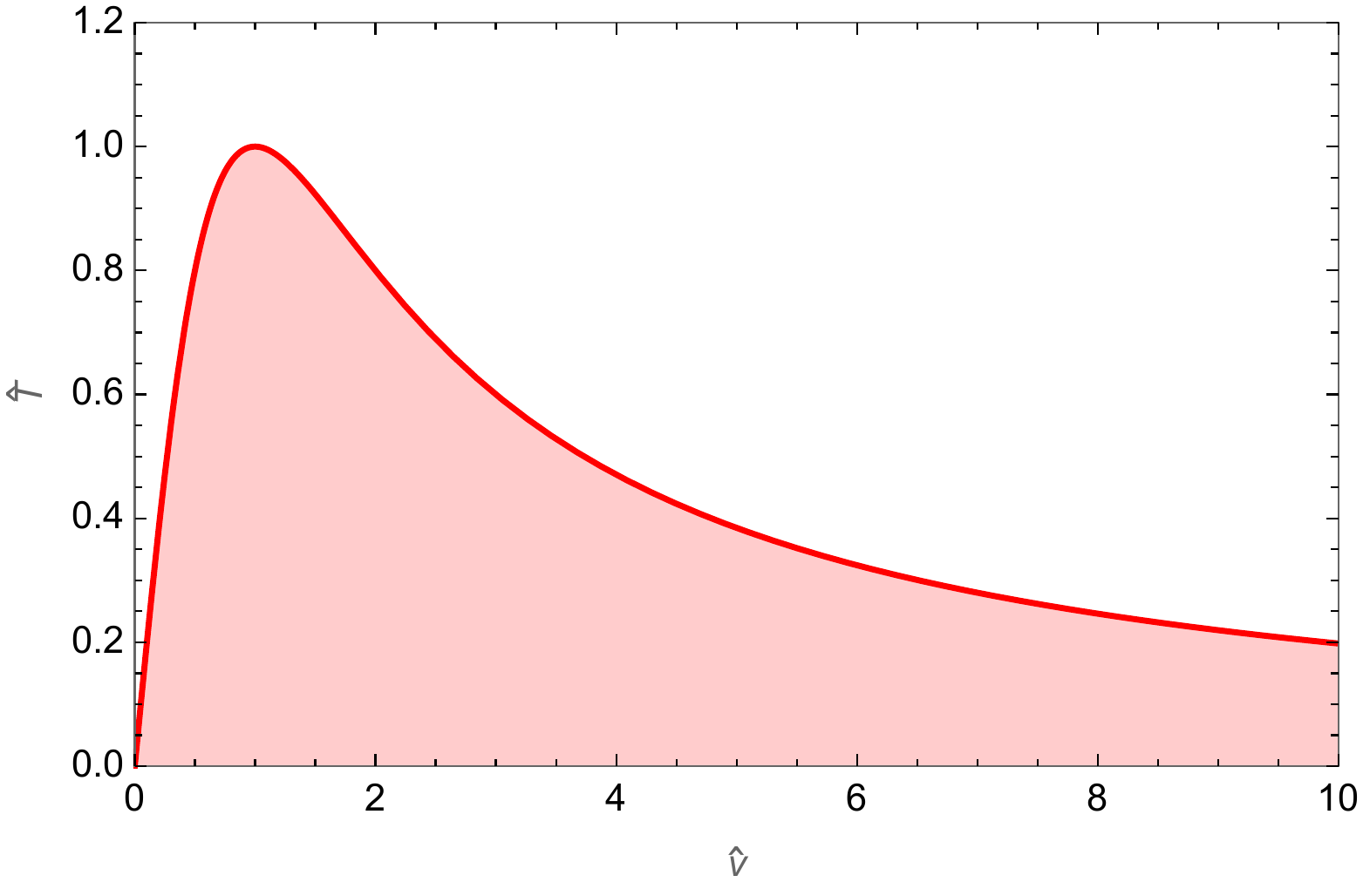}
	\includegraphics[scale=0.5]{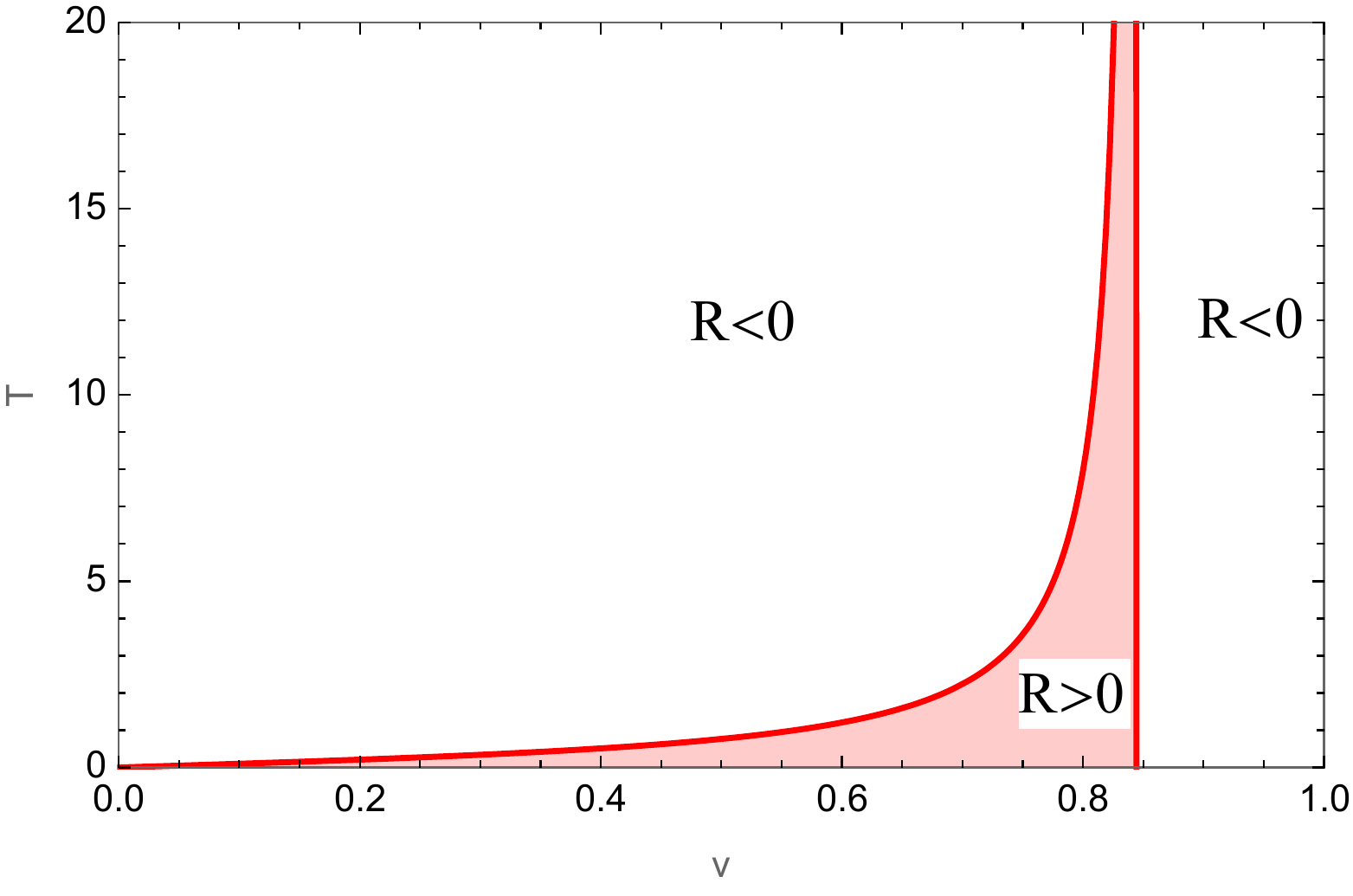}
	\caption{Left: The spinodal curve of the five-dimensional neutral GB-AdS black hole, Right: The sign-changing curve where the scalar
		curvature $R^{in}$ changes its sign for $\alpha=0.2$.\label{SpinGB}}
\end{figure}


\end{document}